\documentclass[a4paper,twoside,twocolumn,british,journal,breaklinks]{IEEEtran}
\usepackage[T1]{fontenc}
\usepackage{babel}
\usepackage{units}
\usepackage{amsmath}
\usepackage{amssymb}
\usepackage{graphicx}
\usepackage{xargs}[2008/03/08]
\usepackage[unicode=true]
 {hyperref}

\makeatletter

\pdfpageheight\paperheight
\pdfpagewidth\paperwidth

 \let\oldforeign@language\foreign@language
 \DeclareRobustCommand{\foreign@language}[1]{%
   \lowercase{\oldforeign@language{#1}}}

\usepackage{pdfpages}
\usepackage{tikz}

\usepackage{pgfplots} 
\pgfplotsset{compat=newest} 
\pgfplotsset{plot coordinates/math parser=false} 
\pgfplotsset{every axis/.append style={font=\footnotesize}}
\pgfplotsset{tick label style={font=\footnotesize}}
\usetikzlibrary{plotmarks}

\newlength\figureheight 
\newlength\figurewidth 
\usepackage{grffile}

\usepackage{enumitem}
\setlist[itemize]{leftmargin=*}

\usepackage{cite}
 
\usepackage[caption=false,font=footnotesize]{subfig}

\newcommand{\example}[1]{#1}

\usepackage[showonlyrefs]{mathtools}

\usepackage{glossaries}
\renewcommand{\newacronym}[4][]{\newglossaryentry{#2}{type=\acronymtype,
name={#3},
short={#3},
long={#4},
shortplural={#3\glspluralsuffix},
longplural={#4\glspluralsuffix},
description={#4},
text={#3},
first={#4~(#3)},
plural={#3\glspluralsuffix},
firstplural={#4\glspluralsuffix~(#3\glspluralsuffix)},
#1}}
\newacronym{BLA}{BLA}{Best Linear Approximation}
\newacronym{MIMO}{MIMO}{Multiple-Input Multiple-Output}
\newacronym{SISO}{SISO}{Single-Input Single-Output}
\newacronym{SIMO}{SIMO}{Single-Input Multiple-Output}
\newacronym{PISPO}{PISPO}{Period-In Same Period-Out}
\newacronym{FRF}{FRF}{Frequency Response Function}
\newacronym{DCA}{DCA}{Distortion Contribution Analysis}
\newacronym{FFT}{FFT}{Fast Fourier Transform}
\newacronym{HB}{HB}{Harmonic Balance}
\newacronym{ADS}{ADS}{Advanced Design System}
\newacronym{LSSS}{LSSS}{Large-Signal Small-Signal}
\newacronym{DFT}{DFT}{Discrete Fourier Transform}
\newacronym{PDF}{PDF}{Probability Density Function}
\newacronym{PSD}{PSD}{Power Spectral Density}
\newacronym{RPM}{RPM}{Random Phase Multisine}
\newacronym{RMS}{rms}{root mean square}
\newacronym{PLL}{PLL}{Phase-Locked Loop}
\newacronym{VCO}{VCO}{Voltage Controlled Oscillator}
\newacronym{PFD}{PFD}{Phase-Frequency Detector}
\newacronym{CP}{CP}{Charge Pump}
\newacronym{OTA}{OTA}{Operational Transconductance Amplifier}
\newacronym{VUB}{VUB}{Vrije Universiteit Brussel}
\makeglossaries

\hypersetup{draft}

\makeatother

\begin{document}

\global\long\def\Fourier{\digamma\!}

\global\long\def\t{\!\left(t\right)}

\global\long\def\dd{d}

\global\long\def\D{D}

\global\long\def\Dvec{\mathbf{D}}

\global\long\def\BLA{^{\mathrm{BLA}}}

\newcommandx\stageNr[2][usedefault, addprefix=\global, 1=n, 2=]{_{\left[#1\right]#2}}

\global\long\def\n{^{\left[n\right]}}

\global\long\def\mm{^{\left(m\right)}}

\global\long\def\DistoMIMO{\D_{t}}

\global\long\def\DistoSISO{\D_{t}}

\global\long\def\wk{\!\left(j\omega_{k}\right)}

\global\long\def\k{\!\left(k\right)}

\global\long\def\vecop{\mathrm{vec}}

\global\long\def\herm{\mathsf{H}}

\title{Distortion Contribution Analysis\\
with the Best Linear Approximation}

\author{Adam~Cooman,~\IEEEmembership{Student~Member,~IEEE}, Piet~Bronders,~\IEEEmembership{Student~Member,~IEEE},
Dries~Peumans,~\IEEEmembership{Student~Member,~IEEE}, Gerd~Vandersteen,~\IEEEmembership{Member,~IEEE}
and~Yves~Rolain,~\IEEEmembership{Fellow,~IEEE}\thanks{Adam~Cooman, Piet Bronders, Dries Peumans, Gerd~Vandersteen and
Yves~Rolain are with the Department ELEC of the VUB, Brussels, Belgium,
e-mail: \protect\href{http://acooman@vub.ac.be}{acooman@vub.ac.be}.}}

\markboth{IEEE TRANSACTIONS ON CIRCUITS AND SYSTEMS\textemdash I: REGULAR PAPERS}{A. Cooman \MakeLowercase{\emph{et al.}}: Distortion Contribution
Analysis with the Best Linear Approximation}

\IEEEpubid{}
\maketitle
\begin{abstract}
A \gls{DCA} obtains the distortion at the output of an analog electronic
circuit as a sum of distortion contributions of its sub-circuits.
Similar to a noise analysis, a \gls{DCA} helps a designer to pinpoint
the actual source of the distortion.

Classically, the \gls{DCA} uses the Volterra theory to model the
circuit and its sub-circuits. This \gls{DCA} has been proven useful
for small circuits or heavily simplified examples. In more complex
circuits however, the amount of contributions increases quickly, making
the interpretation of the results difficult.

In this paper, the \gls{BLA} is used to perform the \gls{DCA} instead.
The \gls{BLA} represents the behaviour of a sub-circuit as a linear
circuit with the unmodelled distortion represented by a noise source.
Combining the \gls{BLA} with a classic noise analysis yields a \gls{DCA}
that is simple to understand, yet capable to handle complex excitation
signals and complex strongly non-linear circuits.
\end{abstract}


\IEEEpeerreviewmaketitle{}

\glsresetall

The decrease of supply voltages in aggressively scaled technologies
results in non-linear distortion to become one of the main limiting
factors for the dynamic range of analog electronic circuits. Still,
the distortion is often taken into account at later stages of the
design only by using a one or two-tone test. The total harmonic distortion
or intermodulation distortion is then intended to describe the non-linearity
of the circuitry. These numbers give an indication of the total distortion
without providing an in depth insight into its origin.

The aim of the \gls{DCA} is to split the total distortion into contributions
of each sub-circuit \cite{Wambacq1998}. Comparing the different contributions
allows the designer to pinpoint the dominant sources of non-linear
distortion and hereby effectively reduce the total distortion \cite{Aikio2011}.
Note that, taken from a birds-eye view, the \gls{DCA} closely resembles
a noise analysis. The difference is that it is now applied to non-linear
distortion sources.

The first \gls{DCA} methods were based on Volterra theory~\cite{Narayanan1967,Wambacq1998}.
The method has been illustrated on small circuits and has extensively
been used in both the analysis and the design of electronic circuits~\cite{Wambacq1998}.
However, the Volterra-based \gls{DCA} has some severe limitations:
\begin{itemize}
\item[-] Only weakly non-linear circuits can be analysed. In strongly non-linear
circuits, the Volterra series obtained around the DC operating point
fail to converge, which limits the use of the \gls{DCA}. An extension
to strongly non-linear circuits has been proposed in \cite{Aikio2005}.
\item[-] Only smaller circuits can be analysed. The number of Volterra distortion
contributions rises quickly for larger circuits. A simple Miller op-amp,
for example, yields over 700 contributions~\cite{Wambacq1998}, making
interpretation of these results more difficult if not impossible.
\item[-] Only the distortion under single-tone or two-tone excitation signals
is considered. Exciting a circuit with practical complex modulated
excitation signals, however, has a big influence on the non-linear
behaviour of the circuit and hence on the distortion~\cite{DeLocht2006}.
\end{itemize}
More recently, the \gls{BLA} has been used to perform a \gls{DCA}
on analog electronic circuits. The idea was originally proposed in~\cite{DeLocht2004}
and has been applied to several examples in the past~\cite{Bos2008,Bronckers2008,Vandersteen2009a,Cooman2012a}.
In the \gls{BLA} framework, the behaviour of a non-linear system
is approximated in least-squares sense by a linear system. As a consequence,
the distortion introduced by the system can be represented by an additive
noise source. Combining the \gls{BLA} analysis with a classic noise
analysis yields a \gls{DCA} which solves some of the drawbacks of
the classic Volterra-based implementations, at the cost of an increased
simulation time. The main benefits of the method are:
\begin{itemize}
\item[-] Linear models are used to describe the dynamic behaviour of the sub-circuits,
while the distortion in the circuit is represented by noise-like sources.
The concept of linear dynamic systems and noise are familiar to all
designers.
\item[-] The analysis also applies to modulated excitation signals, which
leads to an accurate and realistic representation of the non-linear
distortion generated by the circuit in real operation.
\item[-] The \gls{BLA} method does not require simplified device models or
the access to internal nodes of the device models.
\item[-] The validity of the \gls{BLA} is not restricted to weakly non-linear
circuits. Strongly non-linear power amplifiers and hard saturation
can still be modelled with the \gls{BLA}. We rule out strongly non-linear
circuits designed for frequency translation like mixers in this paper
however.
\end{itemize}
All previous implementations of the \gls{BLA}-based \gls{DCA} use
a simplified representation of the circuit, ignoring possible correlation
of the distortion introduced by different stages on the one hand and
input-output impedances of the circuit on the other hand\cite{DeLocht2004,DeLocht2007,Bronckers2008,Vandersteen2009a}.
In this paper, we link the \gls{BLA}-based \gls{DCA} to the theoretical
framework of the \gls{BLA}~\cite{Pintelon2012,Pintelon2013,Pintelon2011}
(Section~\ref{sec:BLAbasics} and~\ref{sec:BLA_DCA}). A first contribution
is to correctly take the correlation between different distortion
sources present in the circuit into account. Secondly, the \gls{BLA}-based
\gls{DCA} is extended to the use of S-parameters to represent the
sub-circuits (Section~\ref{sec:BLA_DCA_port}). This extension takes
reverse gain and terminal impedances of the sub-circuits into account,
which enables a \gls{BLA}-based \gls{DCA} at the transistor level.
The introduction of S-parameters moves the sub-circuit representation
from \gls{SISO} sub-blocks to \gls{MIMO} sub-blocks, which complicates
the identification of the \gls{BLA}. Section~\ref{sec:BLA_estimation}
details the simulations required to estimate the \gls{BLA} of the
\gls{MIMO} sub-circuits correctly. Finally, the \gls{BLA}-based
\gls{DCA} is applied to a two-stage Miller op-amp, a Doherty power
amplifier and a gm-C biquad to show the benefits and general applicability
of the method (Section~\ref{sec:Examples}).

\section{The Single Input Single Output\protect \\
Best Linear Approximation\label{sec:BLAbasics}}

\begin{figure}
\begin{centering}
\includegraphics{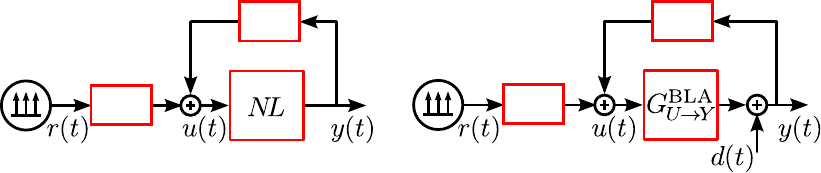}
\par\end{centering}
\caption{\label{fig:BLAbasics}The \gls{BLA} framework allows one to model
non-linear systems as the combination of a linear system $G_{U\rightarrow Y}\protect\BLA$
and a noise-like source $\protect\D$ which represents the non-linear
distortion.}
\end{figure}

Instead of working with deterministic input signals, such as a sine
wave or a two-tone, the \gls{BLA} framework considers noise excitation
signals with a fixed \gls{PSD} and \gls{PDF}. Examples are filtered
white Gaussian noise and telecommunication signals with a specified
bandwidth~\cite{Carvalho2008}. When the excitation signals from
the specified class of signals are applied to a \gls{PISPO} non-linear
system\footnote{The \gls{PISPO} class of non-linear systems includes hard non-linear
elements like saturation or switching, but rules out systems that
generate sub-harmonics, hysteresis or behave chaotically \cite{Pintelon2012}.}, the response of the system can be approximated in least-squares
sense by the \glsentryfirst{BLA}~\cite{Pintelon2012}. Consider
a \glsentryfirst{SISO} non-linear system with input $u\t$ and output
$y\t$ placed in a feedback configuration (Fig. \ref{fig:BLAbasics}).
The whole system is excited by a reference signal $r\t$ with fixed
\gls{PSD} and \gls{PDF}. The \gls{BLA} of the system is then defined
as 
\begin{equation}
G_{U\rightarrow Y}\BLA\!\left(j\omega\right)=\frac{S_{yr}\!\left(j\omega\right)}{S_{ur}\!\left(j\omega\right)}=\frac{\Fourier\left\{ \mathbb{E}\left\{ y\t r\!\left(t-\tau\right)\right\} \right\} }{\Fourier\left\{ \mathbb{E}\left\{ u\t r\!\left(t-\tau\right)\right\} \right\} }\label{eq:BLA_definition}
\end{equation}
where $S_{yr}$ and $S_{ur}$ are the cross-power spectrum between
the reference signal $r\t$ and the output $y\t$ and input $u\t$
respectively \cite{Pintelon2012,Pintelon2013}. $\Fourier\left\{ x\t\right\} $
represents the Fourier transform of $x\t$ and the expected value
operator $\mathbb{E}\left\{ \right\} $ is taken with respect to the
random reference signal $r\t$. 

The difference between the actual output $y\t$ of the non-linear
system and the output predicted by the \gls{BLA} is denoted by $\dd\t$.
The distortion term $\dd\t$ is zero mean, uncorrelated with the reference
signal $r\t$ and behaves like noise \cite{Pintelon2013}. In the
frequency domain, the input-output relation at each excited frequency
bin $k$ is written as\footnote{We use the notation of \cite{Pintelon2012,Pintelon2013}, where frequency
bins in signals are indicated with $\k$ and the corresponding frequencies
in a frequency response are denoted with $\wk$.}:
\begin{equation}
Y\k=G_{U\rightarrow Y}\BLA\wk U\k+\D\k\label{eq:InOutRelation}
\end{equation}
wherein $Y\k$ and $U\k$ are the \gls{DFT} spectra of $y\t$ and
$u\t$ respectively, evaluated at the $k^{\mathrm{th}}$ frequency
bin. Equation~\eqref{eq:InOutRelation}~is the key expression that
allows to use the \gls{BLA} in a \gls{DCA}. It tells that the output
of each sub-circuit can be written as the sum of the output of a signal-dependent
linear dynamic circuit and an additive noise source $\D\k$ representing
the distortion. Calculating the frequency response function from each
distortion source to the considered output of the total circuit allows
to compute the different distortion contributions.

\subsection{Multisine Excitations}

Instead of working with noisy excitation signals directly, \glspl{RPM}
are commonly used to estimate the \gls{BLA} for a Gaussian input
signal. A \gls{RPM} is a sum of harmonically related sine waves with
a random phase:
\begin{equation}
r\t=\sum_{k=1}^{N}A_{k}\sin\left(2\pi kf_{0}t+\phi_{k}\right)\label{eq:multisine}
\end{equation}
where $f_{0}$ is the base frequency of the multisine. $A_{k}$ and
$\phi_{k}$ are the amplitude and phase of the $k^{\mathrm{th}}$
tone in the multisine. When the phases are drawn randomly from a uniform
distribution $\left[0,2\pi\right[$, the multisine \gls{PDF} converges
to a Gaussian \gls{PDF} when considering a large number of frequencies
$N$. The amplitude coefficients $A_{k}$ in the multisine can be
chosen in a deterministic way to set the required \gls{PSD}. In RF
applications, the multisine only excites frequency bins around a centre
frequency $f_{\mathrm{c}}$ between a minimum frequency $f_{\mathrm{min}}$
and a maximum frequency $f_{\mathrm{max}}$. $f_{\mathrm{min}}$,
$f_{\mathrm{max}}$ and $f_{\mathrm{c}}$ are all set to integer multiples
of the base frequency $f_{0}$. In baseband applications, lowpass
multisines are used which excite frequencies starting from DC, so
$f_{\mathrm{min}}$ is equal to $f_{0}$ there.

To separate even and odd non-linear distortion contributions in a
baseband circuit, odd lowpass multisines are commonly used ($A_{k}\!\!=\!\!0$
for even $k$). An even non-linearity always combines an even number
of frequencies in the multisine, so its distortion contributions will
fall on the even frequency bins. An odd non-linearity will only return
contributions on odd frequency bins, so just by design of the excitation
signal, the even and odd non-linear contributions generated in the
circuit are split.

The odd non-linear distortion will end up on the excited frequency
lines, which complicates estimating the amount of odd-order distortion
between $f_{\mathrm{min}}$ and $f_{\mathrm{max}}$. To overcome this
issue Random-Odd \glspl{RPM} are used \cite{Pintelon2004a}. In a
random-odd \gls{RPM}, one odd excited line is left out randomly out
of groups of three. On these ``detection lines\textquotedbl{}, an
estimate of the odd-non-linear distortion is easily obtained. The
use of such a Random-odd \gls{RPM} is illustrated in the first example.

\subsection{Determining the \gls{SISO} \gls{BLA}}

We introduce the ``robust method\textquotedbl{} to determine the
\gls{BLA} of a circuit as it allows to estimate the \gls{BLA} and
the distortion in the circuit with the highest accuracy \cite{Pintelon2004a}.
$M$ different-phase multisines are applied to the system. In those
different-phase multisines only the $\phi_{k}$ are changed in \eqref{eq:multisine},
the amount of tones ($N$) and the amplitude of the tones ($A_{k}$)
is kept the same for each multisine. The steady-state response of
the circuit to each of the different-phase multisines is then determined
using a large-signal simulation\footnote{A Transient simulation, Periodic Steady-State (PSS), Harmonic Balance
(HB) or Envelope simulation can all be used to determine the steady-state
response to the multisine excitation.}. The steady-state spectrum of the reference signal, input signal
and output signal of the circuit under excitation by the $m^{\mathrm{th}}$
different-phase multisine is labelled $R\mm$, $U\mm$ and $Y\mm$
respectively.

The \gls{BLA} of the system is now obtained in a two-step procedure.
First, the \gls{SIMO} \gls{BLA} from the reference signal to the
stacked output-input vector ($\mathbf{Z}$) is determined by averaging
over the steady-state response to the different-phase multisines:
\begin{align}
\mathbf{Z}\mm\wk & =\left[\begin{array}{c}
Y\mm\k\\
U\mm\k
\end{array}\right]\left(R\mm\k\right)^{-1}\nonumber \\
\mathbf{Z}\wk & =\frac{1}{M}\sum_{m=1}^{M}\mathbf{Z}\mm\wk=\left[\begin{array}{c}
G_{R\rightarrow Y}\BLA\wk\\
G_{R\rightarrow U}\BLA\wk
\end{array}\right]\label{eq:Zbla}\\
\mathbf{r}_{\mathbf{Z}}\mm\wk & =\mathbf{Z}\mm\wk-\mathbf{Z}\wk\nonumber \\
\mathbf{C}_{\mathbf{Z}}\wk & =\frac{1}{M\left(M-1\right)}\sum_{m=1}^{M}\mathbf{r}_{\mathbf{Z}}\mm\wk\left[\mathbf{r}_{\mathbf{Z}}\mm\wk\right]^{\herm}\label{eq:Cz}
\end{align}
$\mathbf{C}_{\mathbf{Z}}$ is the sample covariance matrix of $\mathbf{Z}$,
expressing the uncertainty on the estimate. $\cdot^{\herm}$ indicates
the Hermitian transpose. Finally, the \gls{BLA} of the system operating
in feedback is determined as:

\begin{equation}
G_{U\rightarrow Y}\BLA\wk=\frac{G_{R\rightarrow Y}\BLA\wk}{G_{R\rightarrow U}\BLA\wk}\label{eq:Gbla}
\end{equation}
Furthermore, the uncertainty on the \gls{BLA}-estimate can be calculated
as:
\begin{flalign*}
\sigma_{G_{U\rightarrow Y}\BLA}^{2}\wk & =\left|\frac{1}{G_{R\rightarrow U}\BLA\wk}\right|^{2}\mathbf{V}\wk\mathbf{C}_{\mathbf{Z}}\wk\mathbf{V}^{\herm}\wk\\
\mathbf{V}\wk & =\left[\begin{array}{cc}
1 & -G_{U\rightarrow Y}\BLA\wk\end{array}\right]
\end{flalign*}
More details about the experiments and algorithm needed to determine
the \gls{BLA} are given in Section~\ref{sec:BLA_estimation} and
in references~\cite{Pintelon2012,Pintelon2011,Pintelon2013}.

The estimate of the uncertainty on the \gls{BLA} is used to determine
the number of phase realisations that is needed to obtain a sufficiently
certain estimate of the \gls{BLA}. When the uncertainty is too high
for the specific application, more phase realisations are simulated
and added to the set of signals until a sufficiently low uncertainty
is obtained. In strongly non-linear circuits, it can take several
hundreds of phase realisations to obtain a good estimate, as the standard
deviation only decreases with the square-root of the number of realisations.

\newpage{}

\noindent \example{
\subsection*{Example 1: \gls{BLA} of a Miller op-amp in feedback}

Before we use the \gls{BLA} in a \gls{DCA}, let us illustrate how
the \gls{BLA} is used to describe the behaviour of an op-amp placed
in a negative feedback configuration. The op-amp under test is a two-stage
Miller op-amp designed in a commercial $0.18\mathrm{\mu m}$ technology
with a Gain-Bandwidth product of $10\mathrm{MHz}$ for a load capacitance
of $10\mathrm{pF}$.
\begin{center}
\includegraphics{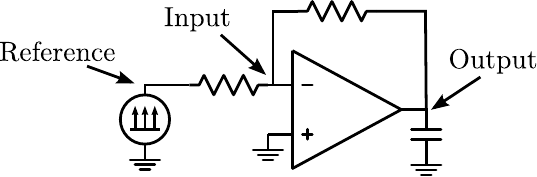}
\par\end{center}
\noindent The reference signals are lowpass random-odd \glspl{RPM}
\eqref{eq:multisine} with $f_{0}\!\!=\!\!f_{\mathrm{min}}\!\!=\!\!0.1\mathrm{kHz}$
and $f_{\mathrm{max}}\!\!=\!\!100\mathrm{kHz}$. The amplitude of
the multisines is chosen flat as a function of frequency and such
that the \gls{RMS} voltage equals $50\mathrm{mV}$. 
\noindent \begin{center}
\includegraphics{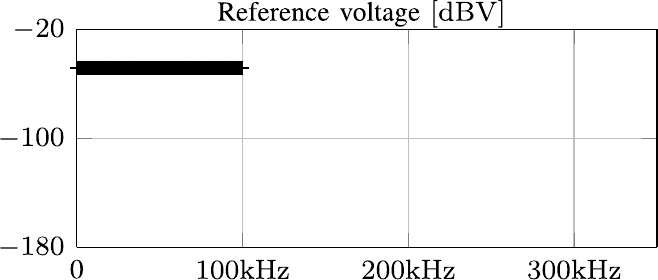}
\par\end{center}
\noindent The input and output voltages obtained with a \gls{HB}
simulation clearly contain non-linear distortion, as there is energy
appearing at non-excited frequency lines. The even frequency bins
are coloured in blue and the non-excited odd frequency bins are indicated
in red. 
\noindent \begin{center}
\includegraphics{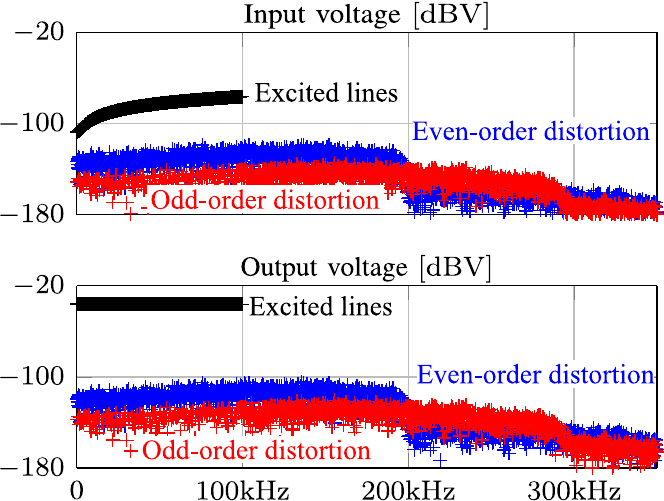}
\par\end{center}
\noindent It is clear that even and odd-order non-linear distortion
are separated by using the odd multisines. The in-band odd non-linear
distortion is visible on the detection lines.

\noindent The \gls{BLA} obtained with $7$ different-phase multisines
is shown below. A compression of $0.1\mathrm{dB}$ is observed with
respect to the results obtained with an AC simulation.
\noindent \begin{center}
\includegraphics{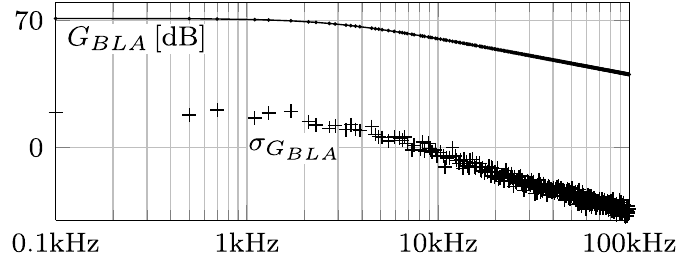}
\par\end{center}}

\newpage{}

\section{Distortion Contribution Analysis \& \gls{BLA}\label{sec:BLA_DCA}}

By fixing the input signal class (fixed \gls{PSD} and fixed \gls{PDF})
and working with the \gls{BLA} framework, the non-linear distortion
in a circuit can be treated as if it were noise. Combining the \gls{BLA}
with a noise analysis then allows one to determine the dominant source
of non-linear distortion in a system. The basic idea is simple (\cite{DeLocht2004,Cooman2013}),
but a rigorous treatment of the concept has not been detailed in literature.
The main difference between the noise analysis in a \gls{BLA}-based
\gls{DCA} and the classic noise analysis is that all distortion sources
are correlated. Taking this correlation into account is very important
to obtain the correct result for the \gls{DCA} and is one of the
main contributions of this paper.

\begin{figure}
\begin{centering}
\includegraphics{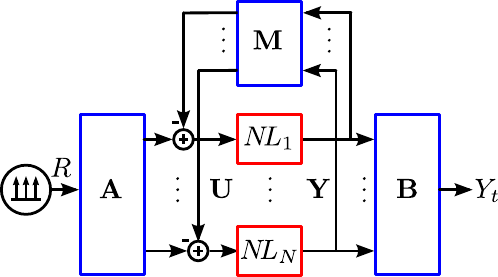}
\par\end{centering}
\caption{\label{fig:problemstatement}The general system under consideration
consists of multiple non-linear systems in a feedback configuration.}
\end{figure}
Consider $N$ \gls{SISO} non-linear systems embedded in a linear
feedback structure as is shown in Fig. \ref{fig:problemstatement}.
The whole system is excited by random-phase multisines $R$ with a
specified \gls{PSD} and \gls{PDF}. Using \eqref{eq:InOutRelation},
the output of the system at the $k^{\mathrm{th}}$ bin of the \gls{DFT}
can be written as
\begin{equation}
Y_{t}\k=G_{R\rightarrow Y_{t}}\BLA\wk R\k+\DistoSISO\k\label{eq:Yt}
\end{equation}
This expression indicates that the output contains a best linear contribution
to the input ($G_{R\rightarrow Y_{t}}\BLA R$) and a distortion term
$\DistoSISO$. The goal of the \gls{DCA} is to write $\DistoSISO$
as a sum of contributions stemming from the $N$ non-linear blocks
in the circuit. $\DistoSISO\k$ has noise-like properties, as was
explained in the previous section. Therefore, only the power of the
output distortion, or $\mathbb{E}\left\{ \DistoSISO\k\DistoSISO^{\herm}\k\right\} $,
can be considered.

To determine the distortion contributions separately, first consider
the \glspl{BLA} of the different non-linear systems. All inputs and
outputs of the non-linear sub-circuits are gathered frequency by frequency
in column vectors $\mathbf{U}\k$ and $\mathbf{Y}\k$:
\begin{equation}
\mathbf{U}\k=\left[\begin{array}{c}
U\stageNr[1]\k\\
\vdots\\
U\stageNr[N]\k
\end{array}\right]\quad\mathbf{Y}\k=\left[\begin{array}{c}
Y\stageNr[1]\k\\
\vdots\\
Y\stageNr[N]\k
\end{array}\right]\label{eq:UYdef}
\end{equation}
where $Y\stageNr\k$ and $U\stageNr\k$ indicate the output and input
\gls{DFT} spectra of the $n^{\mathrm{th}}$ sub-circuit respectively.
The different \gls{SISO} \glspl{BLA}, as defined in \eqref{eq:BLA_definition},
are grouped in a diagonal matrix:
\begin{equation}
\mathbf{G}_{\mathbf{U}\rightarrow\mathbf{Y}}\BLA\wk\!\!=\!\!\left[\!\!\begin{array}{ccc}
G_{U\stageNr[1]\rightarrow Y\stageNr[1]}\BLA\wk\!\! & \!\!\cdots\!\! & 0\\
\vdots & \!\!\ddots\!\! & \vdots\\
0 & \!\!\cdots\!\! & \!\!G_{U\stageNr[N]\rightarrow Y\stageNr[N]}\BLA\wk
\end{array}\!\!\right]\label{eq:BLAdiag}
\end{equation}
The input-output relation of all non-linear systems can now be written
simultaneously as follows:
\begin{equation}
\mathbf{Y}\k=\mathbf{G}_{\mathbf{U}\rightarrow\mathbf{Y}}\BLA\wk\mathbf{U}\k+\Dvec\k\label{eq:BLA_diag}
\end{equation}
Where $\Dvec\in\mathbb{C}^{N\times1}$ contains the non-linear distortion
introduced by the $N$ sub-systems. From here, the frequency indices
$\k$ and $\wk$ will be omitted for notational simplicity. It is
shown in Appendix~\ref{sec:AppendixSISODCA} that the output signal
$Y_{t}$ of the total system can be written as:
\begin{alignat}{1}
Y_{t}= & \overbrace{\mathbf{B}\left(\mathbf{I}_{N}+\mathbf{G}_{\mathbf{U}\rightarrow\mathbf{Y}}\BLA\mathbf{M}\right)^{-1}\mathbf{G}_{\mathbf{U}\rightarrow\mathbf{Y}}\BLA\mathbf{A}}^{G_{R\rightarrow Y_{t}\BLA}}R\label{eq:Yt_asfunctionof_Ys}\\
 & +\underbrace{\mathbf{B}\left(\mathbf{I}_{N}+\mathbf{G}_{\mathbf{U}\rightarrow\mathbf{Y}}\BLA\mathbf{M}\right)^{-1}\Dvec}_{\DistoSISO}\nonumber 
\end{alignat}
Where $\mathbf{B}$, $\mathbf{A}$ and $\mathbf{M}$ are the linear
blocks connected to the non-linear sub-circuits as shown in Fig.~\ref{fig:problemstatement}.
$\mathbf{I}_{N}$ is the identity matrix of size $N$. The second
part of \eqref{eq:Yt_asfunctionof_Ys} yields the expression for the
output distortion as a function of the distortion in the sub-systems.
Considering the power of the distortion at $\k$
\begin{flalign}
\mathbb{E}\left\{ \DistoSISO\DistoSISO^{\herm}\right\} = & \mathbf{T}_{\mathrm{out}}\mathbb{E}\left\{ \Dvec\Dvec^{\herm}\right\} \mathbf{T}_{\mathrm{out}}^{\herm}\label{eq:DCA_formula}
\end{flalign}
with $\mathbf{T}_{\mathrm{out}}=\mathbf{B}\left(\mathbf{I}_{N}+\mathbf{G}_{\mathbf{U}\rightarrow\mathbf{Y}}\BLA\mathbf{M}\right)^{-1}$
a row vector of length $N$ that contains the \gls{FRF} from each
distortion source to the output. The $n^{\mathrm{th}}$ element of
$\mathbf{T}_{\mathrm{out}}$ will be called $T\stageNr[n]$ from now
on. $\mathbb{E}\left\{ \Dvec\Dvec^{\herm}\right\} =\mathbf{C}_{\Dvec}$
is the covariance matrix of the distortion introduced by the non-linear
sub-systems. A two-step procedure is used to obtain an estimate of
$\mathbf{C}_{\Dvec}$, which is similar to the way we determined the
\gls{BLA} itself in section~\ref{sec:BLAbasics} \cite{Pintelon2011}.
First, the covariance matrix of the stacked input-output vectors $\mathbf{C}_{\mathbf{Z}}$
is determined as in equation \eqref{eq:Cz}, but now $Y\mm$ and $U\mm$
are replaced by the stacked input and output signals defined in \eqref{eq:UYdef}.
$\mathbf{C_{\mathbf{Z}}}$ is multiplied by the number of phase realisations
$M$, as we are interested in the power of the distortion, rather
than in the uncertainty on the \gls{BLA}-estimate. This $\mathbf{C}_{\mathbf{Z}}$
is now a $2N\!\times\!2N$ matrix. To obtain a full-rank estimate
of $\mathbf{C}_{\mathbf{Z}}$, the response to at least $2N$ different-phase
multisines must be simulated. $\mathbf{C}_{\Dvec}$ is then calculated
starting from $\mathbf{C}_{\mathbf{Z}}$ in the following way
\begin{equation}
\mathbf{C}_{\Dvec}=M\left[\begin{array}{cc}
\mathbf{I}_{N} & -\mathbf{G}_{\mathbf{U}\rightarrow\mathbf{Y}}\BLA\end{array}\right]\mathbf{C}_{\mathbf{Z}}\left[\begin{array}{cc}
\mathbf{I}_{N} & -\mathbf{G}_{\mathbf{U}\rightarrow\mathbf{Y}}\BLA\end{array}\right]^{\herm}\label{eq:Cys_determination}
\end{equation}
The matrix product in \eqref{eq:DCA_formula} can be re-written as:
\[
\mathbb{E}\left\{ \DistoSISO\DistoSISO^{\herm}\right\} =\sum_{i=1}^{N}\sum_{j=1}^{N}\left[\mathbf{C}_{\Dvec}\right]_{i,j}T\stageNr[i]T\stageNr[j]^{\herm}
\]
Herein $\mathbf{C}_{\Dvec}$ is an Hermitian matrix. The complex conjugate
contributions of $\left[\mathbf{C}_{\Dvec}\right]_{i,j}$ and $\left[\mathbf{C}_{\Dvec}\right]_{j,i}$
will therefore combine to form a single, real-valued distortion power
contribution. The expression for the distortion at the output can
now be simplified as follows:
\begin{flalign}
\mathbb{E}\!\left\{ \!\DistoSISO\DistoSISO^{\herm}\!\right\} = & \sum_{i=1}^{N}\underbrace{\left[\mathbf{C}_{\Dvec}\right]_{i,i}\left|T\stageNr[i]\right|^{2}}_{C_{\stageNr[i]}}+\sum_{i=2}^{N}\sum_{j=1}^{i-1}\underbrace{2\Re\left\{ \left[\mathbf{C}_{\Dvec}\right]_{i,j}T\stageNr[i]T\stageNr[j]^{\herm}\right\} }_{C_{\stageNr[i,j]}}\label{eq:DCA_formula_vec}
\end{flalign}
Equation \eqref{eq:DCA_formula_vec} contains all the different distortion
contributions: each element of the covariance matrix of the distortion
sources is transferred to the output. The total distortion at the
output is then the sum of all these contributions. The contributions
can be sorted according to their magnitude to determine the dominant
distortion contribution.

From now on, we will refer to the distortion contributions due to
the diagonal elements of the distortion covariance matrix as \emph{direct
distortion contributions} ($C_{\stageNr[i]}$). The contributions
due to the off-diagonal elements will be called \emph{correlation
distortion contributions} ($C_{\stageNr[i,j]}$).

\noindent \example{
\subsection*{Example 2: DCA of a non-linearity followed by its inverse}

To clarify the interpretation of the different distortion contributions
obtained with the \gls{BLA}-based \gls{DCA}, we consider the trivial
example of a static non-linearity followed by its inverse. The first
non-linear block is an exponential function and the second its inverse:
a logarithm. The cascade of both blocks results in a perfectly linear
system.
\begin{center}
\includegraphics{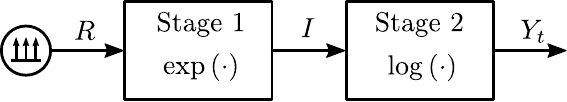}
\par\end{center}
The system is excited by random-phase multisines with an $f_{0}\!=\!1\mathrm{Hz}$
which excite all frequencies up to $100\mathrm{Hz}$. The \gls{RMS}
of the multisine was set to $0.5\mathrm{V}$. The steady-state spectrum
for the signal $I$, measured between the two non-linear blocks, is
shown below.
\noindent \begin{center}
\includegraphics[width=1\columnwidth]{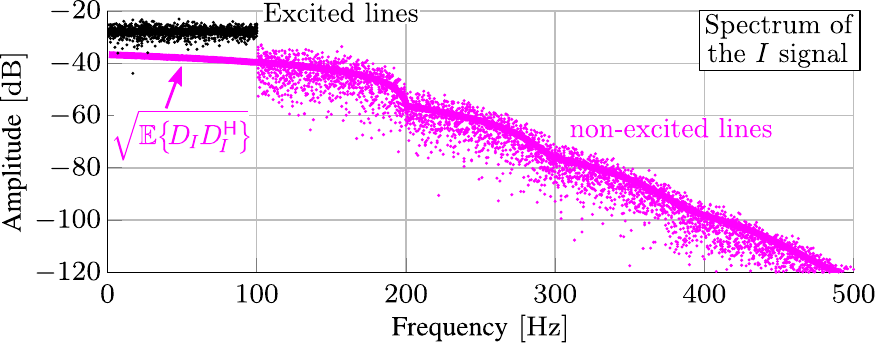}
\par\end{center}
The frequency bins excited by the multisine are shown in black, the
remaining frequency lines in magenta. The measured distortion power
at the internal signal is shown with the magenta line in the plot.
The amount of non-linear distortion at the intermediate signal in
this cascade is very high (signal to distortion ratio of $10\mathrm{dB}$),
but all the distortion is completely cancelled out by the second block,
so that the input and output signals are exactly the same. We calculate
the \gls{BLA} of the two non-linear blocks using \eqref{eq:Zbla}-\eqref{eq:Gbla}.
10000 different-phase multisines were simulated to obtain an adequately
low uncertainty on the \gls{BLA}-estimates in this strongly non-linear
circuit. 

The obtained \glspl{BLA} and their $3\sigma$ uncertainty bound are
shown below:
\noindent \begin{center}
\includegraphics{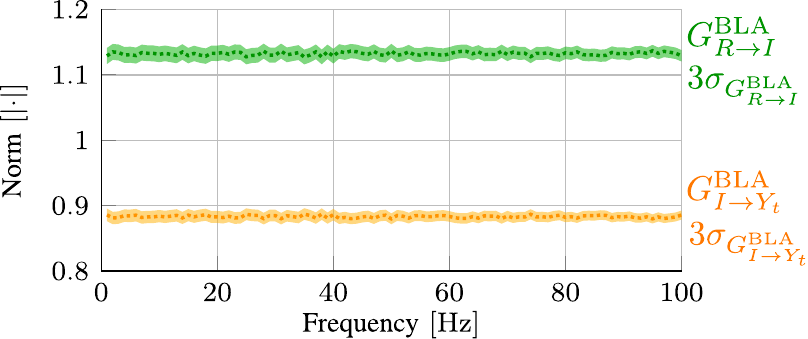}
\par\end{center}
With these \glspl{BLA}, the covariance matrix of the distortion sources
can be calculated using \eqref{eq:Cys_determination}. In this simple
example, there are two distortion sources, one for each non-linearity.
This results in a $2\times2$ covariance matrix $\mathbf{C}_{\mathbf{D}}$.
The elements on the diagonal of $\mathbf{C}_{\mathbf{D}}$ describe
the power of each of the distortion sources. The off-diagonal elements
indicate the correlation between both sources. The values of the distortion
covariance matrix are shown below:
\noindent \begin{center}
\includegraphics{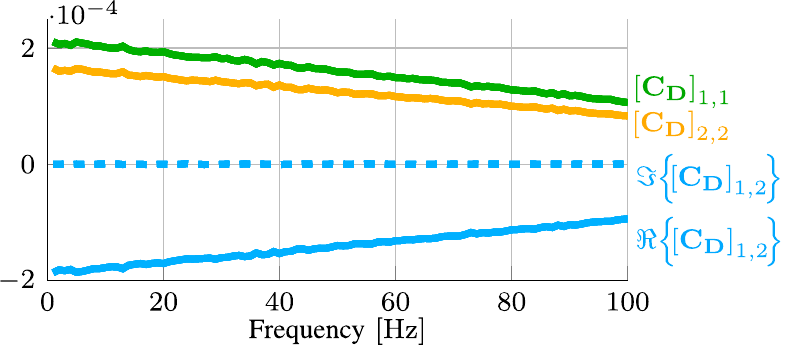}
\par\end{center}
The three distortion contributions to the output can now be calculated.
The \gls{FRF} from each distortion source to the output can be obtained
using \eqref{eq:Yt_asfunctionof_Ys}. For this example, we have:
\[
T\stageNr[1]\wk=G_{I\rightarrow Y_{t}}\BLA\wk\qquad T\stageNr[2]\wk=1
\]
With $\mathbf{C}_{\mathbf{D}}$, $T_{\stageNr[1]}$ and $T_{\stageNr[2]}$,
we can calculate the distortion contributions to the output of the
circuit. The direct distortion contributions due to the first stage
is
\begin{flalign*}
C_{\stageNr[1]}\wk & =\left|G_{R\rightarrow I}\BLA\wk\right|^{2}\left[\mathbf{C}_{\mathbf{D}}\wk\right]_{1,1}
\end{flalign*}
The direct distortion contributions due to the second stage is
\[
C_{\stageNr[1]}\wk=\left[\mathbf{C}_{\mathbf{D}}\wk\right]_{2,2}
\]
The correlation distortion contribution is given by
\[
C_{\stageNr[1,2]}\wk=G_{I\rightarrow Y_{t}}\BLA\wk\left[\mathbf{C}_{\mathbf{D}}\wk\right]_{2,2}
\]
The obtained distortion contributions are plotted below:
\noindent \begin{center}
\includegraphics{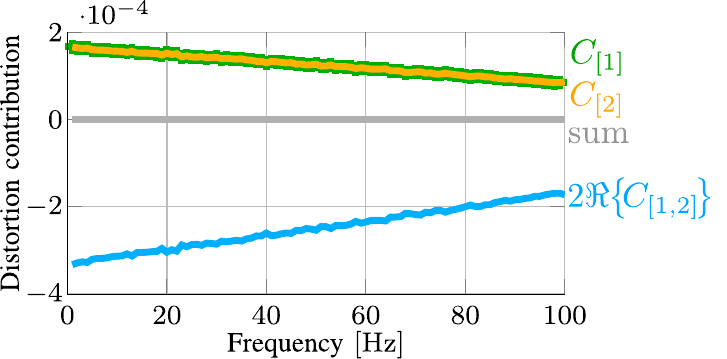}
\par\end{center}
The two direct contributions are equal in amplitude and both positive.
The correlation contribution is equal to the sum of the two direct
contributions, but opposite in sign. The sum of all contributions
(shown in gray on the plot above) therefore lies very close to zero.

With this very simple example we have shown the effectiveness of the
\gls{BLA}-based \gls{DCA} to predict the distortion contributions
of a strongly non-linear circuit under a modulated excitation signal.
Additionally, we have shown that it is important to keep the correlation
distortion contributions into account to obtain a correct result.}

\newpage{}

\section{\gls{BLA}-based \gls{DCA} with S-parameters\label{sec:BLA_DCA_port}}

The previous expressions can be used in a \gls{DCA} on system-level
simulations, where every sub-circuit is represented by a \glsentryfirst{SISO}
system. In actual electronic circuits however, a port-based representation
of the sub-blocks has to be used to represent the terminal impedances
and to include the forward and reverse gain of each sub-circuit in
the circuit. In the remainder of this paper, S-parameters will be
used to represent the behaviour of the different circuit blocks. Similar
expressions can be obtained for the Y and Z parameters, but this is
considered to be outside of the scope of this paper. 

The reasoning in this section is very similar to the one detailed
in the previous section but, instead of working with \gls{SISO} \glspl{BLA},
each of the sub-circuits is described by a \glsentryfirst{MIMO} \gls{BLA}.

\begin{figure}
\begin{centering}
\includegraphics{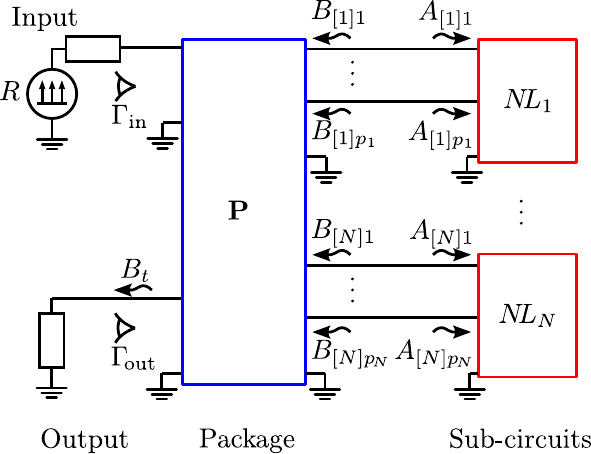}
\par\end{centering}
\caption{\label{fig:CUT}The circuit under test will consist of $N$ non-linear
sub-circuits embedded in a linear package. The whole circuit is excited
by a reference signal $R$. The goal of the \gls{DCA} is split the
distortion in the output wave $B_{t}$ into its contributions.}
\end{figure}

Fig.~\ref{fig:CUT} shows the general circuit under test for the
\gls{DCA}. There are $N$ non-linear sub-circuits embedded in a package.
The whole circuit is excited by different-phase multisines $R$ and
the output of the circuit is terminated in a load impedance. The steady-state
port voltages and currents of the sub-circuits are measured and transformed
into waves using the classical expression~\cite{Kurokawa1965}:
\[
B_{i}=\frac{V_{i}-Z_{0}I_{i}}{2\sqrt{Z_{0}}}\qquad A_{i}=\frac{V_{i}+Z_{0}I_{i}}{2\sqrt{Z_{0}}}
\]
where $V_{i}$ is the port voltage and $I_{i}$ is the port current
flowing into the sub-circuit port. $Z_{0}$ is a user-chosen reference
impedance. The $A$ and $B$ waves at the $p_{n}$ ports of the $n^{\mathrm{th}}$
sub-circuit are gathered in vectors
\begin{flalign*}
\mathbf{B}\stageNr & =\left[\begin{array}{c}
B_{\stageNr[n][1]}\\
\vdots\\
B_{\stageNr[n][p_{n}]}
\end{array}\right]\quad\mathbf{A}\stageNr=\left[\begin{array}{c}
A_{\stageNr[n][1]}\\
\vdots\\
A_{\stageNr[n][p_{n}]}
\end{array}\right]
\end{flalign*}
The relation between $\mathbf{A}\stageNr$ and $\mathbf{B}\stageNr$
is given by the \gls{MIMO} \gls{BLA} $\mathbf{S}_{\mathbf{A}\stageNr\rightarrow\mathbf{B}\stageNr}\BLA$:
\[
\mathbf{B}\stageNr=\mathbf{S}_{\mathbf{A}\stageNr\rightarrow\mathbf{B}\stageNr}\BLA\mathbf{A}\stageNr+\Dvec\stageNr
\]
$\Dvec\stageNr$ is the vector of distortion sources. Determining
the \gls{MIMO} \gls{BLA} is more complex than what has been done
for the \gls{SIMO} procedure described in Section~\ref{sec:BLAbasics}.
The algorithm needed is described in the following section. For now,
assume the \gls{MIMO} \gls{BLA} to be known. All the different \glspl{BLA}
are gathered in a block diagonal matrix, similarly to what was done
in~\eqref{eq:BLA_diag}:
\begin{equation}
\left[\!\!\begin{array}{c}
\mathbf{B}\stageNr[1]\\
\vdots\\
\mathbf{B}\stageNr[N]
\end{array}\!\!\right]\!\!=\!\!\left[\!\!\begin{array}{ccc}
\mathbf{S}_{\mathbf{A}\stageNr[1]\rightarrow\mathbf{B}\stageNr[1]}\BLA\!\! & \!\!\cdots\!\! & \mathbf{0}\\
\vdots & \!\!\ddots\!\! & \vdots\\
\mathbf{0} & \!\!\cdots\!\! & \!\!\mathbf{S}_{\mathbf{A}\stageNr[N]\rightarrow\mathbf{B}\stageNr[N]}\BLA
\end{array}\!\!\right]\!\!\left[\!\!\!\begin{array}{c}
\mathbf{A}\stageNr[1]\\
\vdots\\
\mathbf{A}\stageNr[N]
\end{array}\!\!\!\right]\!\!+\!\!\overbrace{\left[\!\!\begin{array}{c}
\Dvec\stageNr[1]\\
\vdots\\
\Dvec\stageNr[N]
\end{array}\negmedspace\!\right]}^{\mathbf{D}}\label{eq:Sbla}
\end{equation}
The total number of ports of all the sub-circuits is denoted by $P$.
The vector of distortion sources $\Dvec\in\mathbb{C}^{P\times1}$
is again noise-like, so the covariance matrix $\mathbf{C}_{\Dvec}\!=\!\mathbb{E}\left\{ \Dvec\Dvec^{\herm}\right\} $
is used to describe it. Determining $\mathbf{C}_{\Dvec}$ is done
in the same way as explained in Section~\ref{sec:BLA_DCA} and equation~\eqref{eq:Cys_determination}.

The distortion at the output of the system is defined by considering
the \gls{BLA} from the reference multisine to the output wave $B_{t}$:
\begin{equation}
B_{t}=G_{R\rightarrow B_{t}}\BLA R+\DistoMIMO\label{eq:Sbla_total}
\end{equation}
The goal of the \gls{DCA} is to split the power in $\DistoMIMO$
as a sum of contributions from $\mathbf{C}_{\Dvec}$. Classic papers
on wave-based circuit and noise analysis have dealt with this problem
already~(\cite{Monaco1974,Dobrowolski1989}) and describe how to
determine a row vector $\mathbf{T}_{\mathrm{out}}$ that is used to
refer $\mathbf{C}_{\Dvec}$ to the output wave (See appendix~\ref{sec:AppendixMIMODCA}).
\begin{equation}
\mathbb{E}\left\{ \DistoMIMO\DistoMIMO^{\herm}\right\} =\mathbf{T}_{\mathrm{out}}\mathbf{C}_{\Dvec}\mathbf{T}_{\mathrm{out}}^{\herm}\label{eq:DCA_eq_circuit}
\end{equation}
This expression can be re-written in a similar way as in~\eqref{eq:DCA_formula_vec}
to obtain a list of direct distortion contributions and correlation
distortion contributions.

\subsection{Dealing with the combinatorial explosion\label{subsec:Combining-Contributions}}

In circuits, each port of each sub-circuit will create several distortion
contributions: a single direct contribution and some correlation contributions.
The number of contributions can therefore rise quickly, especially
in fully differential, complex, circuits. In a circuit with $P$ ports,
there are $\nicefrac{1}{2}P\left(P-1\right)$ distortion contributions
to the output.

If the amount of contributions is too large to be easily tractable
and interpretable, the different contributions of a single sub-circuit
can be combined into a single contribution by simply summing the contributions
of each of the ports of one sub-circuit. The co-variances can also
be combined in the same way. Combining the contributions of each stage
reduces the amount of contributions for $N$ sub-circuits to $\nicefrac{1}{2}N\left(N-1\right)$.
If this amount of contributions is still too large for an easy interpretation
of the results, the contributions of several sub-circuits can be combined
into one contribution of a larger sub-circuit. It is a clear advantage
that the \gls{BLA}-based \gls{DCA} can easily be applied hierarchically
as this allows to zoom in selectively on the most contributing parts
of the circuit, while leaving the other sub-circuits aggregated at
a higher level of abstraction.

\newpage{}

\section{Estimating the \gls{MIMO} \gls{BLA} of sub-circuits\label{sec:BLA_estimation}}

In the previous section, it was assumed that the \glsentryfirst{BLA}
of the sub-circuits was known. Determining the \gls{BLA} is the most
difficult and time-consuming step in the \gls{BLA}-based \gls{DCA},
as it requires averaging the large-signal steady-state response of
the circuit over many different-phase multisines. For \gls{SISO}
sub-circuits, the estimation steps detailed in equations~\eqref{eq:Zbla}-\eqref{eq:Gbla}
can be used. In case of \gls{MIMO} sub-circuits, extra excitation
signals need to be added to the circuit to determine the \gls{BLA}.

\subsection{\gls{MIMO} identification in feedback}

The \gls{MIMO} \gls{BLA} of the $n^{\mathrm{th}}$ sub-circuit with
$p_{n}$ inputs and outputs, is defined as an extension of \eqref{eq:Gbla}:
\begin{equation}
\mathbf{S}_{\mathbf{A}\stageNr\rightarrow\mathbf{B}\stageNr}\BLA\wk=\mathbf{G}_{\mathbf{R}\rightarrow\mathbf{B}\stageNr}\BLA\wk\left[\mathbf{G}_{\mathbf{R}\rightarrow\mathbf{A}\stageNr}\BLA\wk\right]^{-1}\label{eq:MIMOBLAdef}
\end{equation}
where $\mathbf{G}_{\mathbf{R}\rightarrow\mathbf{B}\stageNr}\BLA\in\mathbb{C}^{p_{n}\times n_{r}}$
is the \gls{MIMO} \gls{BLA} taken from the $n_{r}$ reference signals
to the output waves of the $n^{\mathrm{th}}$ sub-circuit. $\mathbf{G}_{\mathbf{R}\rightarrow\mathbf{A}\stageNr}\BLA\in\mathbb{C}^{p_{n}\times n_{r}}$
is the \gls{MIMO} \gls{BLA} from the reference signals to the input
waves of the sub-circuit. Since $\mathbf{G}_{\mathbf{R}\rightarrow\mathbf{A}\stageNr}\BLA$
must be invertible, at least $p_{n}$ independent reference signals
have to be present in the circuit. Most circuits are only excited
by one main reference signal, so the required extra reference signals
have to be added artificially to the circuit to allow estimation of
the \gls{MIMO} \gls{BLA} of the sub-circuits.

The extra multisines have to be very small in amplitude with respect
to the large signal to avoid changing the \gls{BLA} of the circuit,
which is why they are called tickler multisines~\cite{DeLocht2007}.
If the ticklers are placed on the same frequency grid as the main
multisines, their response will be overwhelmed by the circuit response
to the main multisines. To avoid this overlap, the frequency grid
of the tickler is shifted by a frequency $\left|f_{\epsilon}\right|\in\left]0,\nicefrac{f_{0}}{2}\right[$.
This technique is called 'Zippering'~\cite{Vanhoenacker2000}. The
zippered ticker multisines are defined by:
\[
r_{\mathrm{tickle}}\left(t\right)=\sum_{h=1}^{N}A_{h}\sin\left(2\pi\left(hf_{0}+f_{\epsilon}\right)t+\phi_{h}\right)
\]
the phases of the tickler multisine ($\phi_{h}$) are drawn randomly
from $\left[0,2\pi\right[$. These tickler multisines will create
contributions on frequencies that are always $f_{\epsilon}$ away
from the spectral lines of the main multisines, so the responses of
the main and tickler multisines are easily separated by looking at
the correct frequency bins. Because their amplitude is very small,
the tickler signals can either be applied one-by-one, while the main
multisine remains active, or applied all simultaneously since each
of the ticklers can be given an independent $f_{\epsilon}$.

To obtain the \gls{BLA} with the zippered multisines, the \gls{SIMO}
\gls{BLA} from each reference signal to the stacked output-input
vector of the sub-circuit is determined first by simple averaging
as in \eqref{eq:Zbla}. The \gls{SIMO} \gls{BLA} from reference
signal $R_{r}$ to the stacked output-input vector will be denoted
$\mathbf{G}_{R_{r}\rightarrow\mathbf{Z}}\BLA$, while its uncertainty
is expressed by the covariance matrix $\mathbf{C}_{\mathbf{G}_{R_{r}\rightarrow\mathbf{Z}}\BLA}$.
The $\mathbf{G}_{R_{r}\rightarrow\mathbf{Z}}\BLA$ are known at the
frequencies of the spectral lines of multisines $R_{r}$. All $\mathbf{G}_{R_{r}\rightarrow\mathbf{Z}\stageNr}\BLA$
are then linearly interpolated to the spectral lines of the main multisines
and gathered in a large matrix:
\begin{equation}
\mathbf{G}_{\mathbf{R}\rightarrow\mathbf{Z}\stageNr}\BLA\!\!=\!\!\left[\!\!\begin{array}{c}
\mathbf{G}_{\mathbf{R}\rightarrow\mathbf{B}\stageNr}\BLA\\
\mathbf{G}_{\mathbf{R}\rightarrow\mathbf{A}\stageNr}\BLA
\end{array}\!\!\right]=\left[\!\!\begin{array}{ccc}
\mathbf{G}_{R_{1}\rightarrow\mathbf{Z}\stageNr}\BLA & \!\!\cdots & \!\!\mathbf{G}_{R_{n_{r}}\rightarrow\mathbf{Z}\stageNr}\BLA\end{array}\right]\label{eq:Z_mimoBLA}
\end{equation}
The \gls{BLA} is now obtained using expression~\eqref{eq:MIMOBLAdef}.
To obtain the uncertainty on the \gls{BLA}, the uncertainty on the
$\mathbf{G}_{\mathbf{Z}}$-matrix is transformed by the following
expression~\cite{Pintelon2011}:
\begin{multline*}
\mathbf{C}_{\vecop\left(\mathbf{S}_{\mathbf{A}\stageNr\rightarrow\mathbf{B}\stageNr}\BLA\right)}=\mathbf{T}\mathbf{C}_{\vecop\left(\mathbf{G}_{\mathbf{R}\rightarrow\mathbf{Z}\stageNr}\BLA\right)}\mathbf{T}^{\herm}\\
\mathrm{with}\,\,\mathbf{T}=\left[\left(\mathbf{G}_{\mathbf{R}\rightarrow\mathbf{A}\stageNr}\BLA\right)^{-1}\right]^{T}\otimes\left[\begin{array}{cc}
\mathbf{I}_{n_{y}} & -\mathbf{S}_{\mathbf{A}\stageNr\rightarrow\mathbf{B}\stageNr}\BLA\end{array}\right]
\end{multline*}
where $\otimes$ indicates the Kronecker product \cite{Brewer1978}.
The covariance matrix $\mathbf{C}_{\vecop\left(\mathbf{G}_{\mathbf{Z}}\right)}$
contains the different covariance matrices of the \gls{SIMO} \glspl{BLA}
on its diagonal:
\[
\mathbf{C}_{\vecop\left(\mathbf{G}_{\mathbf{R}\rightarrow\mathbf{Z}\stageNr}\BLA\right)}=\left[\begin{array}{ccc}
\mathbf{C}_{\mathbf{G}_{R_{1}\rightarrow\mathbf{Z}\stageNr}\BLA} & \cdots & \mathbf{0}\\
\vdots & \ddots & \vdots\\
\mathbf{0} & \cdots & \mathbf{C}_{\mathbf{G}_{R_{n_{r}}\rightarrow\mathbf{Z}\stageNr}\BLA}
\end{array}\right]
\]

Determining the \gls{MIMO} \gls{BLA} with the method described here
can be very time-consuming because many large-signal simulations have
to be run. A speed-up can be obtained using advanced non-parametric
estimation techniques like the local polynomial method \cite{Pintelon2010}.
Alternatively, rational approximations can be estimated to reduce
the noisiness of the obtained \gls{BLA} estimates. 

It is more easy to work with tickler current sources, as they can
be added to a circuit's netlist without introducing extra nodes. Choosing
the best nodes to place the tickler multisines and determining the
amplitude of the ticklers will usually require some user intervention.
When the \gls{BLA} from two reference signals to the the input and
output waves of the circuit are too similar, the $\mathbf{G}_{\mathbf{R}\rightarrow\mathbf{A}\stageNr}\BLA$
matrix will be badly conditioned and its inverse will be difficult
to compute. To obtain a good conditioning of the $\mathbf{G}_{\mathbf{R}\rightarrow\mathbf{A}\stageNr}\BLA$,
it is recommended to connect the tickler sources to nodes that are
close to the ports of the circuit under test. 

The amplitude of the tickler multisines should be small, but the circuit's
response to the tickler signal should lie above the numerical noise
floor at the input and output waves. Setting the correct amplitude
of the tickler signals is therefore done by increasing the amplitude
of the tickler until its response is clearly visible in the steady-state
spectra of all input and output signals of the sub-circuits. 

In circuits with very small reverse gains, obtaining a good estimate
of the reverse gain can be very difficult. For the small reverse gain,
the small-signal behaviour of the sub-circuit is used if it cannot
be estimated reliably~\cite{Cooman2016}.

\vfill{}

\pagebreak{}

\noindent \example{
\subsection*{Example 3: \gls{MIMO} \gls{BLA} of a class-C amplifier}

To demonstrate how the \gls{MIMO} \gls{BLA} of a circuit is estimated,
consider this simple class-C amplifier.
\begin{center}
\includegraphics{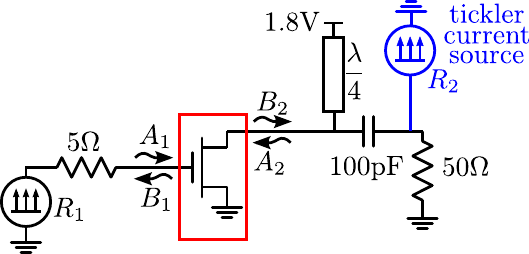}
\par\end{center}
\vspace{-3mm}

\noindent The class-C amplifier is excited by \glspl{RPM} $R_{1}$
that excite $41$ frequencies in a band of $40\mathrm{MHz}$ around
$1\mathrm{GHz}$. Their \gls{RMS} value is $0.2\mathrm{V}$. The
transistor in the amplifier is placed in a common-source configuration,
so it will be modelled by a two-port. There's only one reference multisine
in the circuit so a second multisine current source is added at the
output ($R_{2}$ shown in blue in the figure). The tickler multisines
are shifted $1\mathrm{Hz}$ away from the frequency grid of the main
multisine and are given an \gls{RMS} current of $40\mathrm{\mu A}$.
The resulting $B_{2}$ wave, obtained with \gls{HB} is shown below
\begin{center}
\includegraphics{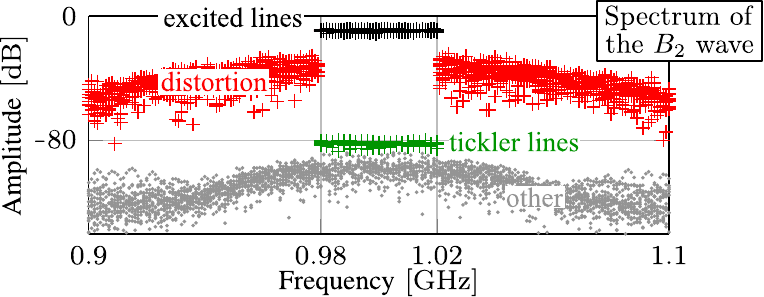}
\par\end{center}
\vspace{-3mm}

\noindent The frequency bins of the main multisines are shown in black,
while its distortion is shown with red. In between the frequency bins
of the main multisines, the response to the tickler is visible. The
frequency bins of the tickler are indicated with green, while all
remaining bins are grey. The amplitude of the tickler was chosen such
that the tickler response was clearly visible above the numeric noise
floor.

The obtained \glspl{BLA} and their $3\sigma$ uncertainty interval
are shown below:
\begin{center}
\includegraphics{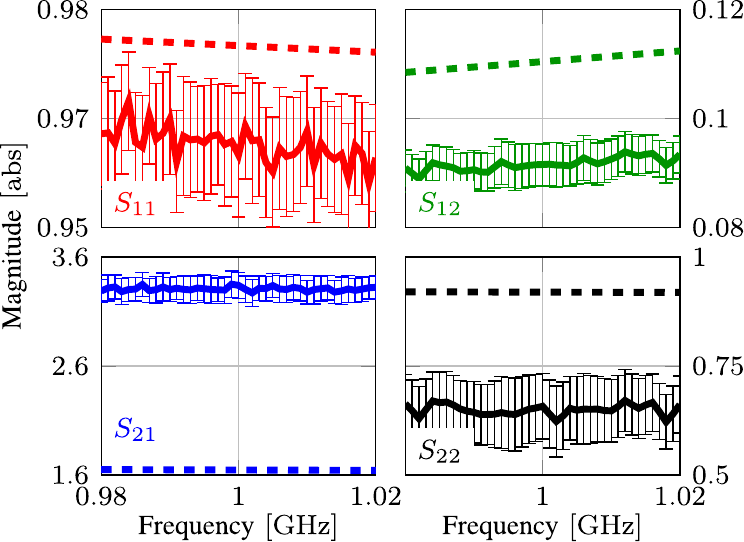}
\par\end{center}
\noindent The dashed lines in the above plots are the small-signal
S-parameters. The largest differences are observed on $S_{21}$ and
$S_{22}$, which is to be expected for a class-C biased transistor.}

\newpage{}

\subsection{Weakly non-linear circuits\label{subsec:Weakly-non-linear-circuits}}

In weakly non-linear circuits, the \gls{BLA} doesn't deviate a lot
from the small-signal frequency response of the circuits, so the small-signal
behaviour can be used to replace the \gls{BLA}. When the \gls{MIMO}
identification scheme is avoided, a significant reduction in simulation
time for the \gls{DCA} is obtained. The small-signal S-parameters
can be obtained quickly in modern simulators combining several AC
or S-parameter simulations.

\noindent Combining small-signal and large-signal results is trivial
when the response of the circuit to the multisine excitations is obtained
with \gls{HB}. If the response is obtained with a time-domain simulation,
the frequency warping should be taken into account properly:
\begin{itemize}
\item[-] A trapezoidal integration method should always be used to avoid artificial
damping of the circuit poles \cite{Kundert1995}.
\item[-] A fixed time step $T_{s}$ should be used to allow calculating the
spectrum easily with the \gls{DFT}.
\item[-] The remaining frequency warping introduced by going to the discrete
time domain should be taken into account. 
\end{itemize}
When the trapezoidal integration method is used, each frequency bin
of the large-signal simulation $kf_{0}$ is warped to a frequency
$f_{\mathrm{warp},k}$ according to the following relationship
\begin{equation}
f_{\mathrm{warp},k}=\frac{1}{\pi T_{s}}\tan\left(\pi kf_{0}T_{s}\right)
\end{equation}
It is as if the circuit is working on the frequency grid determined
by $f_{\mathrm{warp},k}$. Hence, the small-signal behaviour should
be determined on the frequency grid $f_{\mathrm{warp},k}$, or should
be interpolated to the warped frequency grid to minimise errors.

When the \gls{BLA} deviates too far from the small-signal S-parameters
due to strongly non-linear behaviour of the circuit, one cannot replace
the \gls{BLA} by small-signal S-parameters without introducing errors.
It is therefore important to assess the quality of the small-signal
behaviour when using it to predict the distortion in the circuit.
The perfect assessment could be obtained by comparing the small-signal
behaviour to the estimated \gls{MIMO} \gls{BLA} of each sub-circuit,
but calculating the \gls{MIMO} \gls{BLA} beats the purpose of using
the small-signal behaviour in the first place. Instead, the \gls{BLA}
from the reference signal to the input and output waves ($\mathbf{G}_{R_{1}\rightarrow\mathbf{Z}\stageNr}\BLA$
as defined in \eqref{eq:Z_mimoBLA}) can be compared to the frequency
responses obtained with an AC simulation. When the difference between
$\mathbf{G}_{R_{1}\rightarrow\mathbf{Z}\stageNr}\BLA$ and the AC
result lies significantly above the distortion level, the small-signal
behaviour deviates too far from the correct \gls{BLA} and should
not be used in the \gls{DCA}.

\begin{figure}
\noindent \begin{centering}
\includegraphics{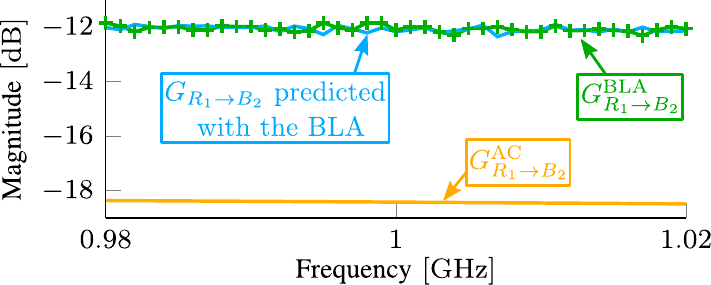}
\par\end{centering}
\caption{\label{fig:SStest}If the small-signal behaviour deviates too far
from the \gls{BLA}, the predicted \gls{BLA} from the main reference
to the output waves is incorrect. The correct \gls{BLA} from reference
to the output wave is shown with (green +). Its
prediction using the small-signal behaviour is shown with (orange -)
. When the \gls{MIMO} \gls{BLA} is used (cyan -), a correct
prediction is obtained.}
\end{figure}

As an example, this test is applied to the class-C amplifier from
before. The \gls{BLA} from the main reference signal to the output
wave is calculated in three different ways: First, the \gls{BLA}
from the reference to the output wave is calculated using the \gls{SISO}
techniques described in Section~\ref{sec:BLAbasics}. Second, the
small-signal frequency response from the reference to the output wave
was calculated with an AC simulation. Finally, the \gls{MIMO} \gls{BLA}
of the transistor was used to predict the same frequency response\footnote{Appendix~\ref{sec:AppendixPredictSIMO} explains how this is done}.
The three frequency responses are shown in Figure~\ref{fig:SStest}.
The frequency response obtained with the AC simulation deviates strongly
from the \gls{BLA} from the reference to the output wave. When the
estimated \gls{MIMO} \gls{BLA} is used to predict the frequency
response, the correct result is obtained.

In this class-C amplifier, it is therefore not possible to use the
small-signal S-parameters in the \gls{DCA}, which is to be expected
for such a strongly non-linear circuit. In the examples shown later,
two circuits are shown where this small-signal assumption is valid.

\section{Examples\label{sec:Examples}}

The \gls{BLA}-based \glsentryfirst{DCA} will now be applied to several
examples. First, a two-stage Miller op-amp is analysed to show the
importance of the correlation between distortion sources in a real
examples. With the second example, a Doherty power amplifier, the
\gls{DCA} is shown to work for strongly non-linear circuits as well.
The weakly non-linear assumption doesn't hold in such an amplifier,
so the \gls{BLA} of the sub-circuits is estimated with zippered multisines.
The third and last example deals with a considerably larger circuit.
The distortion contribution analysis of a a fully differential Gm-C
biquad is used to illustrate how the \gls{DCA} can be applied hierarchically.

\subsection{Miller Op-amp}

\begin{figure}
\begin{centering}
\includegraphics{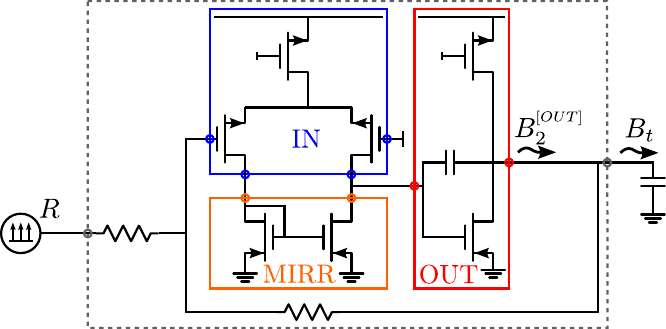}
\par\end{centering}
\caption{\label{fig:example_MILLER_circuit}The two-stage Miller op-amp under
test.}

\vspace{0.1cm}

\noindent \begin{centering}
\includegraphics{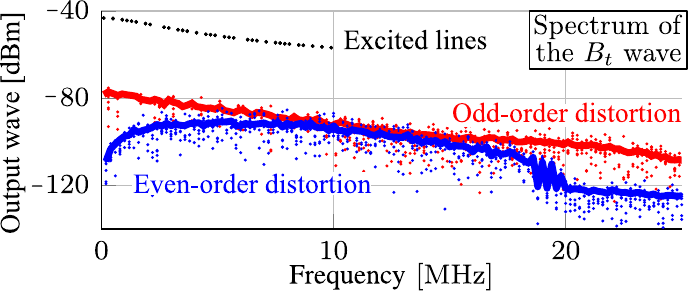}
\par\end{centering}
\vspace{-0.3cm}

\caption{\label{fig:example_MILLER_spectra}Output spectrum of the Miller op-amp
obtained with $6$ different-phase multisines. The excited bins are
shown in black, the even bins in blue and the non-excited odd bins
in red. The \gls{RMS} of the distortion is shown with a blue and
red line for the even and odd bins respectively.}

\begin{centering}
\includegraphics{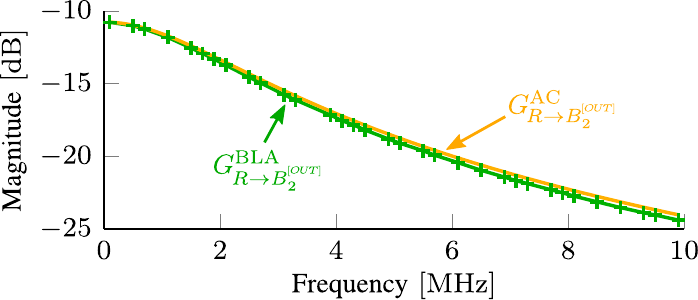}
\par\end{centering}
\caption{\label{fig:example_MILLER_BLAtest}The frequency response from input
voltage to the output wave of the circuit (orange -) doesn't
lie far from the corresponding \gls{BLA} (green +),
so the small-signal S-parameters can be used to represent the sub-circuits
instead of the \gls{MIMO} \gls{BLA}.}

\vspace{0.2cm}

\begin{centering}
\includegraphics[width=1\columnwidth]{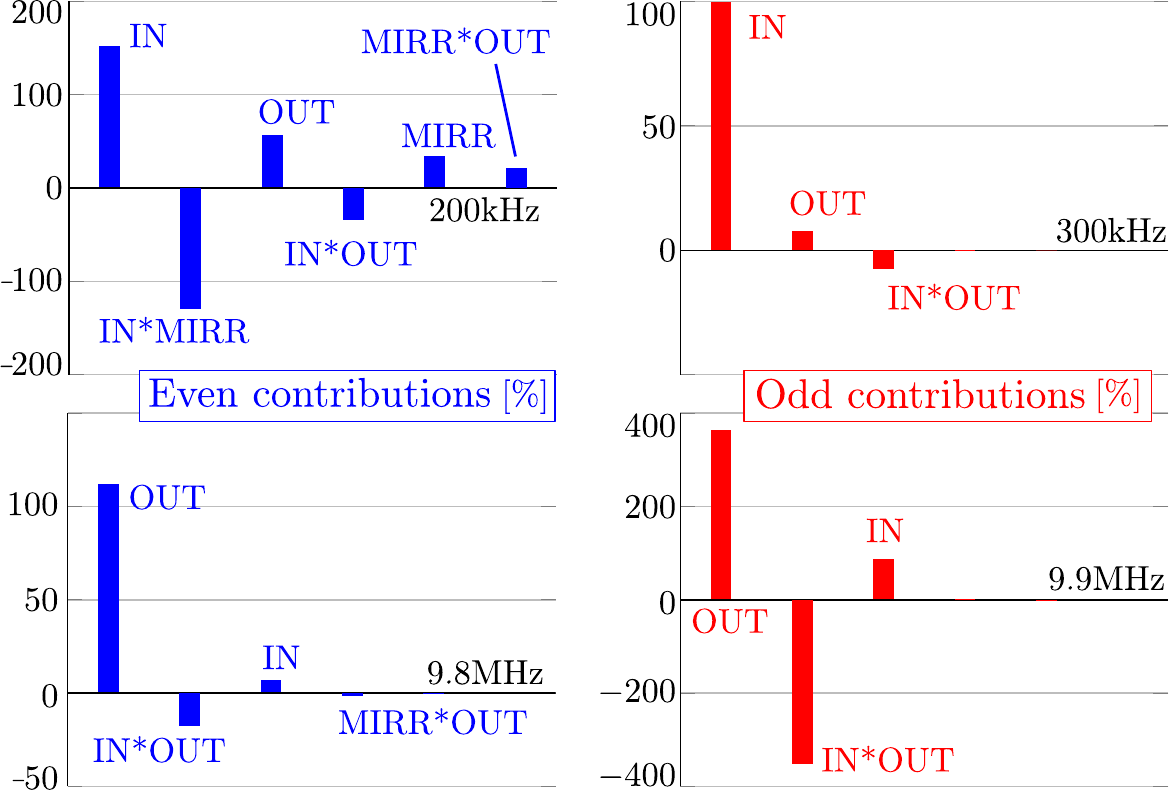}
\par\end{centering}
\caption{\label{fig:example_MILLER_disto}Distortion contributions to the output
wave of the Miller op-amp at four different frequencies}
\end{figure}

As a first example of a \gls{DCA} on the circuit level, we consider
a two-stage Miller-compensated op-amp designed in a commercial $0.18\mathrm{\mu m}$
CMOS technology (Fig.~\ref{fig:example_MILLER_circuit}). The op-amp
is placed in an inverting feedback configuration with a gain of $5$
and drives a load capacitance of $10\mathrm{pF}$, resulting in a
gain-bandwidth product of $10\mathrm{MHz}$. The circuit is split
into three sub-circuits: the input stage, which has three ports, the
current mirror, and the output stage which both have two ports.

The amplifier is excited by lowpass random-odd multisines with a base
frequency $f_{0}$ of $100\mathrm{kHz}$ and $f_{\mathrm{max}}\!\!=\!\!10\mathrm{MHz}$.
The \gls{RMS} voltage of the multisines is set to $0.1\mathrm{V}$.
The steady-state response of the circuit to $50$ different-phase
multisines is obtained by \gls{HB} simulation. The obtained spectrum
of the output wave $B_{t}$ is shown in Fig.~\ref{fig:example_MILLER_spectra}.

The Miller op-amp can be considered to be weakly non-linear, so the
small-signal S-parameters were used to represent each sub-circuit.
To test the validity of this small-signal assumption, the \gls{BLA}
from the reference multisine to all waves in the circuit was compared
to the result obtained with an AC simulation as explained in Section~\eqref{subsec:Weakly-non-linear-circuits}.
The largest difference is observed on the frequency response from
the reference to the output wave of the second stage and it is shown
in Figure~\eqref{fig:example_MILLER_BLAtest}. This difference is
small enough to allow using the small-signal S-parameters instead
of the \gls{MIMO} \gls{BLA} to represent each sub-circuit.

The results of the \glsentryfirst{DCA} at four different frequencies
are shown in Fig.~\ref{fig:example_MILLER_disto}. The contributions
are combined for each stage as was explained in Section~\ref{subsec:Combining-Contributions}.
For three sub-circuits, this results in six distortion contributions,
three direct distortion contributions (one for each stage) and three
correlation distortion contributions.

The even and odd distortion contributions can be split again because
odd \glspl{RPM} were used. At $200\mathrm{kHz}$ (Top left in figure~\ref{fig:example_MILLER_disto})
we obtain only even-order non-linear contributions. The input stage
seems to generate the most distortion with a direct contribution which
is $150\%$ of the total output distortion at that frequency bin,
but its contribution is largely cancelled in its correlation with
the current mirror, which has a contribution of $-120\%$ of the total
distortion. This leads to the output stage as the dominant source
of distortion at $200\mathrm{kHz}$. This shows the importance of
the correlation between the distortion sources in the circuit.

In the odd-order contributions at $300\mathrm{kHz}$ (Top right in
figure~\ref{fig:example_MILLER_disto}), the input stage is clearly
the dominant generator of non-linear distortion.

The bottom series of plots in Fig.~\ref{fig:example_MILLER_disto}
shows the results at the high-frequency end of the analysed band.
At $9.8\mathrm{MHz}$, the output stage dominates the even-order contributions.
At $9.9\mathrm{MHz}$, the odd contribution of the output stage ($350\%$)
is compensated with a co-variance with the input stage ($-350\%$),
making the input stage the dominant source of distortion at this frequency.

With this example, we have shown that the \gls{BLA}-based \gls{DCA}
can be used in circuit-level simulations to obtain the distortion
contributions. Taking the correlation between distortion sources of
different sub-circuits into account is crucial to obtain an accurate
representation of the distortion in this circuit.

\subsection{Doherty Power Amplifier}

\begin{figure}
\begin{centering}
\includegraphics[width=1\columnwidth]{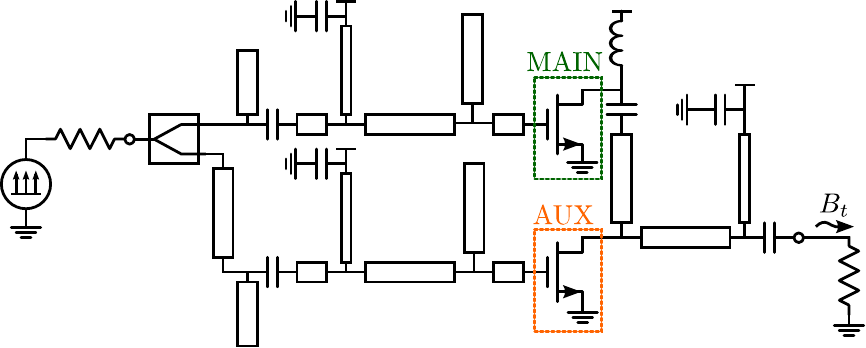}
\par\end{centering}
\caption{\label{fig:example_DOHERTY_circuit}The Doherty power amplifier.}

\vspace{0.1cm}

\noindent \begin{centering}
\includegraphics{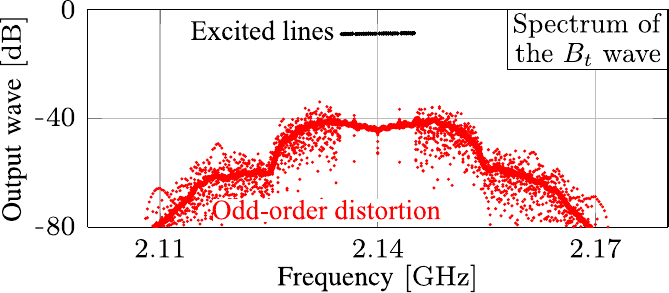}
\par\end{centering}
\vspace{-0.3cm}

\caption{\label{fig:example_DOHERTY_spectra}The output spectrum of the Doherty
power amplifier. The excited lines are shown with black dots.
In a band-pass system only odd non-linear distortion (red dots)
falls inside the band. The \gls{RMS} of the distortion is shown with
(magenta -).}

\vspace{0.2cm}

\noindent \begin{centering}
\includegraphics{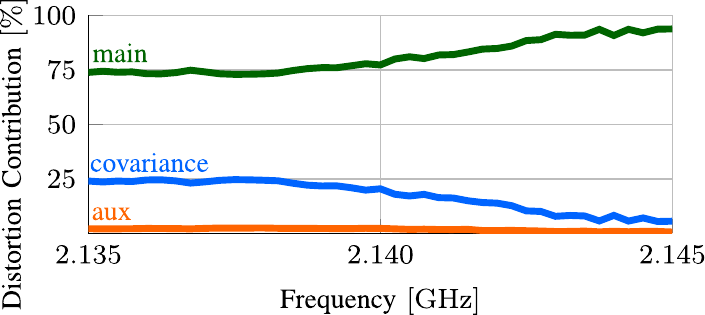}
\par\end{centering}
\vspace{-0.3cm}

\caption{\label{fig:example_DOHERTY_disto}The distortion contributions show
that the main amplifier is the dominant source of non-linear distortion
in the Doherty amplifier.}
\end{figure}

The second example that will be considered is a Doherty power amplifier
found in the example library of Keysight's \gls{ADS}. The amplifier
is built with two Freescale MRF8S21100H transistors for a centre frequency
of $2.14\mathrm{GHz}$ (Fig.~\ref{fig:example_DOHERTY_circuit}).
The main transistor is biased in class-AB with a quiescent current
of $0.7\mathrm{A}$. The auxiliary transistor is biased deep in class-C
with a quiescent current of $1\mathrm{mA}$.

The amplifier is excited by bandpass multisines that have $41$ spectral
lines in a band of $10\mathrm{MHz}$ around $2.14\mathrm{GHz}$. The
\gls{RMS} of the input multisines is $22\mathrm{dBm}$. The steady-state
response of the circuit to the different-phase multisine excitations
is obtained by \gls{HB} simulation. The output wave around the centre
frequency is shown in Fig.~\ref{fig:example_DOHERTY_spectra}. 

This circuit can be considered to be strongly non-linear, especially
because of the auxiliary amplifier. To obtain the \gls{BLA} of the
two transistors in the circuit, tickler multisines are added to the
circuit at the output of the total amplifier. The added multisines
are current sources which insert an \gls{RMS} current of $1\mathrm{\mu A}$
on frequency bins in between the frequencies of the main multisine.

The main distortion contributor is found to be the main transistor
(Fig.~\ref{fig:example_DOHERTY_disto}). This can be expected, as
the auxiliary amplifier only kicks in for limited amounts of time
in this Doherty configuration. A similar Doherty amplifier was analysed
in \cite{Aikio2012} with a Volterra-based \gls{DCA} under two-tone
excitation\footnote{We don't have access to the circuit simulated in \cite{Aikio2012},
nor to their \gls{DCA} method, so only a qualitative comparison between
both methods can be obtained.}. It was concluded there that the auxiliary amplifier only contributes
significantly to the distortion for very high amplitudes in the two-tone.
With modulated signals, like the multisines used in the \gls{BLA}-based
\gls{DCA}, the peaks only occur from time to time, so the average
contribution of the auxiliary amplifier to the total distortion is
low.

The information given by the \gls{BLA}-based \gls{DCA} is limited
for the Doherty power amplifier, because only the signals around the
carrier frequency are used here. Designers are also interested in
how the low-frequency signals in the bias network are up-converted
in-band through the second-order non-linearities \cite{Aikio2011}.
The current implementation of the \gls{DCA} doesn't split the up-converted
low-frequency signals from the high-frequency odd-order non-linear
distortion appearing in-band. A more advanced \gls{BLA}-based \gls{DCA}
can be implemented using the higher-order \glspl{BLA} \cite{Thorsell2011b}.
With the higher-order \gls{BLA}, one could obtain a similar result
to \cite{Aikio2012}, but for modulated signals, instead of two-tones. 

\subsection{Gm-C filter}

\begin{figure}
\begin{centering}
\includegraphics[width=1\columnwidth]{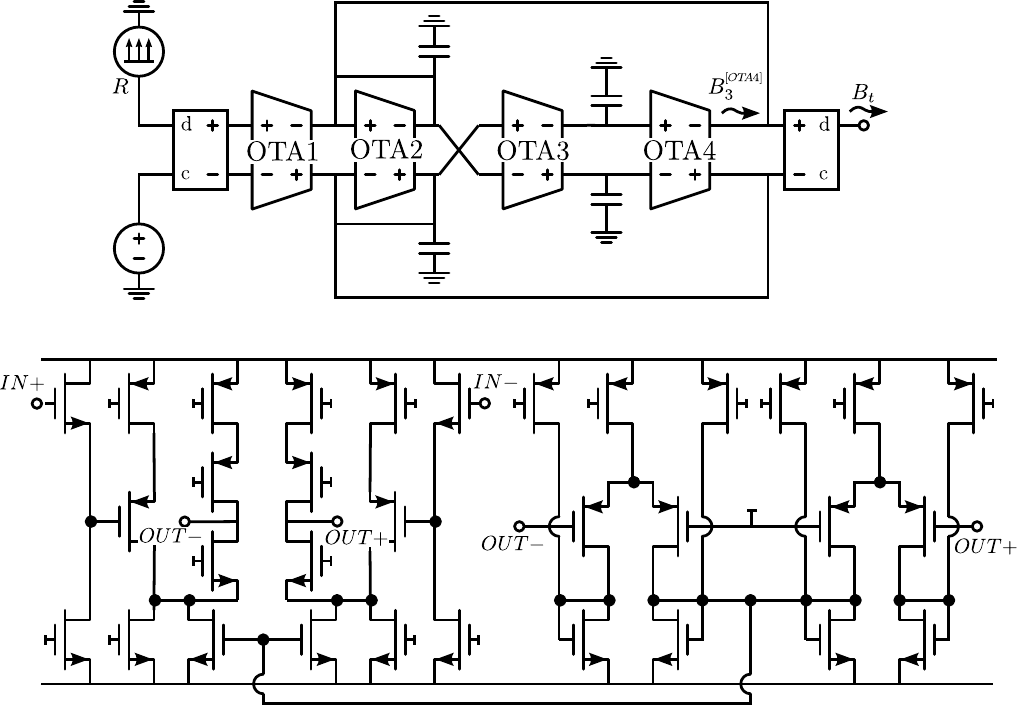}
\par\end{centering}
\caption{\label{fig:example_GmC_circuit}The fully-differential gm-C biquad
under test (top) consists of four identical \glspl{OTA} (bottom).
Only the differential mode of the whole circuit is excited with a
multisines through a balun.}

\vspace{0.2cm}

\noindent \begin{centering}
\includegraphics{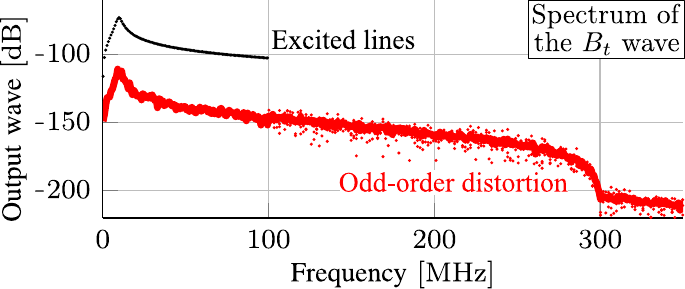}
\par\end{centering}
\vspace{-0.3cm}

\caption{\label{fig:example_GmC_spectra}The spectrum of the output wave shows
only odd non-linear distortion (red dots), as can be expected
in a differential circuit. The excited frequency lines are indicated
with (black dots).The \gls{RMS} of the distortion is shown
with a red line.}

\setlength{\figurewidth}{0.7\columnwidth} 
\setlength{\figureheight}{0.25\columnwidth} 
\begin{centering}
\includegraphics{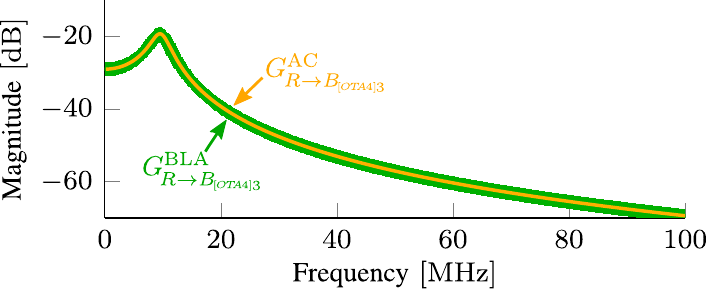}
\par\end{centering}
\vspace{-0.3cm}

\caption{\label{fig:example_GmC_BLAcheck}The small-signal frequency response
from input voltage to one of the output waves of OTA4 (orange -)
doesn't lie far from the corresponding \gls{BLA} (green +),
so the small-signal S-parameters can be used to represent the sub-circuits
instead of the \gls{MIMO} \gls{BLA}.}

\vspace{0.2cm}

\begin{centering}
\includegraphics{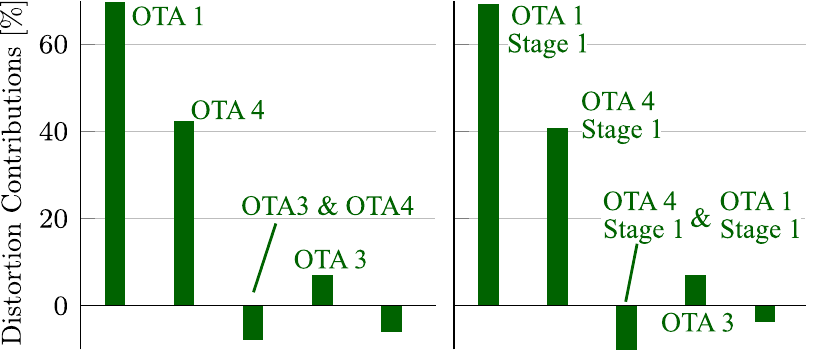}
\par\end{centering}
\caption{\label{fig:example_GmC_disto}The first \gls{OTA} is found to be
the dominant source of distortion in the resonant peak of the circuit
(left). The contribution of the fourth \gls{OTA} cannot be ignored
however. By applying the \gls{DCA} hierarchically (right), it is
found that the first stage is the dominant contributor in both \glspl{OTA}. }
\end{figure}

The final example is a fully differential Gm-C biquad \cite{Kardontchik1992}
designed in the same commercial $0.18\mathrm{\mu m}$ CMOS technology
as the other examples (Fig.~\ref{fig:example_GmC_circuit}). Each
\gls{OTA} in the biquad consists of an input pair and a cascode stage.
The common-mode feedback in the \gls{OTA} is active. The biquad is
configured to create a resonant pole pair at $10\mathrm{MHz}$.

The differential mode of the biquad is excited by full lowpass \glspl{RPM}
(all frequency lines are excited). The multisines have $f_{0}\!\!=\!\!f_{\mathrm{min}}\!\!=\!\!200\mathrm{kHz}$
and $f_{\mathrm{max}}\!\!=\!\!100\mathrm{MHz}$. In a resonant system
like this, the frequency resolution of the multisines should be chosen
to have several lines in the resonance~\cite{Geerardyn2013}. If,
for example, only a single spectral line is placed in a sharp resonance,
the \gls{PDF} of the internal signals will tend to that of a sine
wave, instead of the wanted Gaussian \gls{PDF}. The wanted noise-like
properties of the internal signals in the circuit then disappear,
which is unwanted if the results are to be valid for Gaussian input
signals. The \gls{RMS} of the multisines was set to $50\mathrm{mV}$
and the steady-state response of the circuit to $50$ different-phase
multisines was obtained with \gls{HB}. The resulting spectrum at
the differential output is shown in Fig.~\ref{fig:example_GmC_spectra}.
The distortion at the output lies $50\mathrm{dB}$ below the signal
level, so the circuit is behaving close to linear. No even-order contributions
are present due to the differential nature and perfect symmetry in
the simulations of the circuit. Note that the obtained odd-order distortion
at the output shows a strong frequency dependence around the resonance.

The sub-circuits in this biquad are assumed to be weakly non-linear,
so the 4-port S-parameters of each \gls{OTA} were used in the \gls{DCA}.
This small-signal assumption was verified by comparing the frequency
response from the input of the total circuit to each of the waves
in the circuit with the corresponding \glspl{BLA}. The largest difference
was observed on the frequency response from the reference to the output
waves of OTA4 (shown in Figure~\eqref{fig:example_GmC_BLAcheck}),
but this difference is small enough to consider the small-signal assumption
to be valid.

The first \gls{OTA} is found to be the dominant source of distortion
in the resonance peak of this circuit (Fig.~\ref{fig:example_GmC_disto},
left). The fourth \gls{OTA} also introduces a considerable contribution.
To find out which part of the \gls{OTA} is mainly responsible, the
first and fourth \gls{OTA} were split into two parts and the \gls{DCA}
was applied again. With this hierarchical application of the \gls{BLA}-based
\gls{DCA}, it is found that the first stage of both \glspl{OTA}
1 and 4 are the dominant contribution (Fig.~\ref{fig:example_GmC_disto}
right).

With this final example, we have demonstrated how the \gls{BLA}-based
\gls{DCA} can be used in larger circuits and how it can be used hierarchically
to zoom in on certain sub-circuits to determine the actual source
of non-linear distortion.

\newpage{}

\section{Conclusions}

Combining the \glsentryfirst{BLA} with a noise analysis creates a
\glsentryfirst{DCA} that pin-points the sources of non-linear distortion
in a circuit under excitation of complex, modulated signals. The method
doesn't require special models and can be implemented easily with
the help of a few post-processing steps on top of a commercial simulator.

The \gls{BLA}-based \gls{DCA} works for both weakly and strongly
non-linear circuits, returns a single distortion contribution for
each sub-circuit and takes the correlation between the distortion
introduced by the sub-circuits into account. The method can be applied
hierarchically and therefore suitable to be applied to large circuits
without introducing hundreds of contributions.

The method has been demonstrated on several examples and is shown
to be able to provide information about the source of non-linear distortion
in a two-stage op-amp, a Doherty power amplifier and a fully differential
gm-C filter.

\section*{{\footnotesize{}Acknowlegment}}

{\footnotesize{}This work is sponsored by the Vrije Universiteit Brussel,
dept. ELEC, Pleinlaan 2, 1050 Brussels, Belgium, Fund for Scientific
Research (FWO-Vlaanderen), Institute for the Promotion of Innovation
through Science and Technology in Flanders (IWT-Vlaanderen), the Flemish
Government (Methusalem), the Belgian Federal Government (IUAP VII)
and the Strategic Research Program of the VUB (SRP-19)}{\footnotesize \par}

\appendices{}

\section{Obtaining expression \eqref{eq:Yt_asfunctionof_Ys}\label{sec:AppendixSISODCA}}

In the circuit shown in Figure~\ref{fig:problemstatement}, the following
equations describe the behaviour of the input, output and feedback
dynamics:
\begin{flalign}
\mathbf{U} & =\mathbf{A}R-\mathbf{M}\mathbf{Y}\label{eq:Ufeedback}\\
Y_{t} & =\mathbf{B}\mathbf{Y}\label{eq:outputDynamics}
\end{flalign}
The relation between $\mathbf{U}$ and $\mathbf{Y}$ is given by the
\gls{BLA}
\begin{equation}
\mathbf{Y}=\mathbf{G}_{\mathbf{U}\rightarrow\mathbf{Y}}\BLA\mathbf{U}+\mathbf{D}\label{eq:BLAappendix}
\end{equation}
Plugging equation \eqref{eq:Ufeedback} into \eqref{eq:BLAappendix}
and solving for $\mathbf{Y}$, we obtain
\begin{flalign*}
\mathbf{Y} & =\left(\mathbf{I}_{N}+\mathbf{G}_{\mathbf{U}\rightarrow\mathbf{Y}}\BLA\mathbf{M}\right)^{-1}\left(\mathbf{G}_{\mathbf{U}\rightarrow\mathbf{Y}}\BLA\mathbf{A}R+\mathbf{D}\right)
\end{flalign*}
Using this expression in \eqref{eq:outputDynamics} and grouping the
terms in $R$ and $\mathbf{D}$ yields equation \eqref{eq:Yt_asfunctionof_Ys}.

\section{\gls{DCA} with S-parameters\label{sec:AppendixMIMODCA}}

\noindent To calculate the contributions of the distortion sources
to the output of the circuit, an algorithm similar to the one described
in \cite{Dobrowolski1989} is used. The different S-matrices of the
components in the circuit shown in Fig.~\ref{fig:CUT} are gathered
in a matrix \textbf{$\mathbf{T}$}, while the distortion sources are
gathered in a vector $\mathbf{N}$:
\[
\mathbf{T}=\left[\begin{array}{cccc}
\Gamma_{\mathrm{in}}\\
 & \Gamma_{\mathrm{out}}\\
 &  & \mathbf{P}\\
 &  &  & \mathbf{S}_{\mathbf{A}\rightarrow\mathbf{B}}\BLA
\end{array}\right]\quad\mathbf{N}=\left[\begin{array}{c}
0\\
0\\
\mathbf{0}_{P+2\times1}\\
\Dvec
\end{array}\right]
\]
where $\Gamma_{\mathrm{in}}$ and $\Gamma_{\mathrm{out}}$ are the
reflection factors presented to the circuit by the reference source
and load respectively. \textbf{$\mathbf{S}_{\mathbf{A}\rightarrow\mathbf{B}}\BLA$}
is the block diagonal matrix of size $P\!\times\!P$ which contains
the \glspl{BLA} of the circuits as defined in equation \eqref{eq:Sbla}.
$\Dvec$ is the vector of distortion sources of length $P$ defined
in the same expression. $\mathbf{P}$ is the S-matrix of the package
defined in Figure~\ref{fig:CUT} of size $\left(P+2\right)\!\times\!\left(P+2\right)$
for a circuit with $2$ external ports. 

The interconnection between the different parts of the circuit is
represented by the following matrix:
\[
\mathbf{C}=\left[\begin{array}{cc}
\begin{array}{cc}
\mathbf{0}_{2\times2} & \mathbf{I}_{2\times2}\\
\mathbf{I}_{2\times2} & \mathbf{0}_{2\times2}
\end{array} & \mathbf{0}_{4\times P}\\
\mathbf{0}_{P\times4} & \begin{array}{cc}
\mathbf{0}_{P\times P} & \mathbf{I}_{P\times P}\\
\mathbf{I}_{P\times P} & \mathbf{0}_{P\times P}
\end{array}
\end{array}\right]
\]
The incident-waves at all ports generated by the sources in $\mathbf{N}$
is given by the following expression
\begin{flalign}
\mathbf{A}_{\mathrm{all}}= & \left(\mathbf{C}-\mathbf{T}\right)^{-1}\mathbf{N}=\mathbf{W}^{-1}\mathbf{N}\label{eq:Aall}
\end{flalign}
Since we are only interested in the wave incident to the load, just
the second row of $\mathbf{W}^{-1}$ is used. Also, the first $P+4$
elements of $\mathbf{W}^{-1}$ can be ignored, because the first $P+4$
elements of $\mathbf{N}$ are zero. This finally leads to the expression
for $\mathbf{T}_{\mathrm{out}}$ used in equation \eqref{eq:DCA_eq_circuit}
\begin{equation}
\mathbf{T}_{\mathrm{out}}=\left[\mathbf{W}^{-1}\right]_{2,P+5..2P+4}
\end{equation}

\section{Predicting the \gls{BLA} from reference to the waves in the circuit\label{sec:AppendixPredictSIMO}}

Predicting the frequency response from the main reference signal to
the input and output waves at the ports of the sub-circuits is done
using the same matrix $\mathbf{W}^{-1}$ as was used in Appendix~\ref{sec:AppendixMIMODCA}.
but now, the $\mathbf{N}$-vector is set to the following:
\[
\mathbf{N}=\left[\begin{array}{c}
\frac{1-\Gamma_{S}}{2\sqrt{Z_{0}}}\\
\mathbf{0}_{2P+1\times1}
\end{array}\right]
\]
where $Z_{0}$ is the chosen reference impedance and $\Gamma_{S}$
is the reflection factor presented by the reference source. All A-waves
in the circuit are now predicted by \eqref{eq:Aall}. The frequency
response from the reference voltage source to the A-waves radiating
into the sub-circuits are found at $\left[\mathbf{A}_{\mathrm{all}}\right]_{4+P+1..4+2P,1}$.
The frequency response from the reference voltage source to the B-waves
at the ports of the sub-circuits are found at $\left[\mathbf{A}_{\mathrm{all}}\right]_{5..5+P,1}$.

\bibliographystyle{IEEEtran}
\bibliography{27C__Users_Adam_Google_Drive_PaperDatabase}

\begin{IEEEbiography}[{\includegraphics[width=1in]{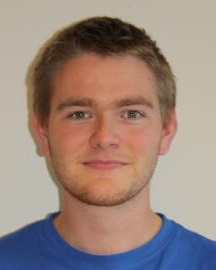}}]{Adam Cooman}
 (1989, Belgium) graduated as an Electrical Engineer in Electronics
and Information Technology in July 2012 at \gls{VUB}. In August 2012
he joined the department ELEC as a PhD student. His main interests
are the design of analog/RF circuits and non-linear modelling.
\end{IEEEbiography}

\begin{IEEEbiography}[{\includegraphics[width=1in]{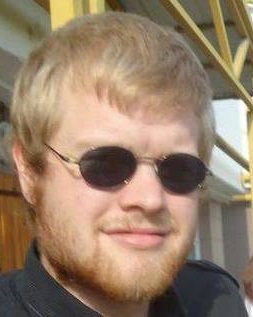}}]{Piet Bronders}
 (1991, Fiji) graduated as an Electrical Engineer in Electronics
and Information Technology in 2014 at the \gls{VUB}. In October 2014
he joined the department ELEC as a PhD student. His main interest
is NL RF circuit design and measurement. His current research topic
is the design and modelling of RF power amplifiers.
\end{IEEEbiography}

\begin{IEEEbiography}[{\includegraphics[width=1in]{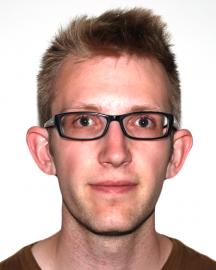}}]{Dries Peumans}
 (1992, Belgium) graduated as an Engineer in Electronics and Information
Technology in 2015 at the \gls{VUB}. Afterwards he joined the department
ELEC as a PhD researcher. His research focuses on non-linear modelling
and the design of analog circuits.
\end{IEEEbiography}

\begin{IEEEbiography}[{\includegraphics[width=1in]{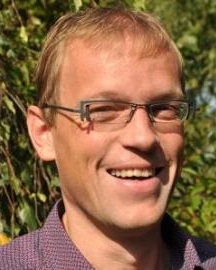}}]{Gerd Vandersteen}
 (1968, Belgium) received the degree in electrical engineering in
1991 and his PhD in electrical engineering in 1997, both from the
\gls{VUB}, Brussels, Belgium. During his postdoc, he worked at the
micro-electronics research centre IMEC as Principal Scientist in the
Wireless Group with the focus on modelling, measurement and simulation
of electronic circuits in state-of-the-art silicon technologies. This
research was in the context of a collaboration with the \gls{VUB}.
From 2008 on, he is working as Prof. at the \gls{VUB} in the department
ELEC within the context of measuring, modelling and analysis of complex
linear and non-linear system. Within this context, the set of systems
under consideration is extended from micro-electronic circuits towards
to all kinds of electro-mechanical systems. From 2011 on, he is director
of the Doctoral School of Natural Sciences and (Bioscience) Engineering
(NSE) at the \gls{VUB}
\end{IEEEbiography}

\begin{IEEEbiography}[{\includegraphics[width=1in]{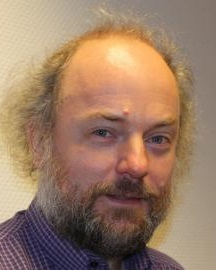}}]{Yves Rolain}
 (1961, Belgium) received the Electrical Engineering (Burgerlijk
Ingenieur) degree in July 1984, the degree of computer sciences in
1986, and the PhD degree in applied sciences in 1993, all from the
\gls{VUB}, Brussels, Belgium. He is currently a research professor
at the VUB in the department ELEC. He became a fellow of the IEEE
in 2006 and received the IEEE I\&M Society award in 2005. His main
interests are microwave measurements and modelling, applied digital
signal processing and system identification.
\end{IEEEbiography}

\end{document}